\documentclass[prc,twocolumn,showpacs,preprintnumbers,amsmath,amssymb,floatfix]{revtex4-1}

\usepackage{tipa}
\usepackage{graphicx}
\usepackage{dcolumn}
\usepackage{bm}
\usepackage{amssymb}
\usepackage{multirow}
\usepackage{booktabs}
\usepackage[section]{placeins}
\usepackage{textcomp}

\begin{document}

\title{$^{25}$Si $\beta^+$-decay spectroscopy}

\date{\today}
\thanks{These authors contributed equally to this work and should be considered co-first authors.\label{Coauthors}}%
\author{L.~J.~Sun$^{1,2}$,\textsuperscript{\ref{Coauthors}}}
\email{sunlijie@msu.edu}
\author{M.~Friedman$^{1,3}$,\textsuperscript{\ref{Coauthors}}}
\email{moshe.friedman@mail.huji.ac.il}
\author{T.~Budner$^{1,4}$}
\author{D.~P\'{e}rez-Loureiro$^{5}$}
\author{E.~Pollacco$^{6}$}
\author{C.~Wrede$^{1,4}$}
\email{wrede@nscl.msu.edu}
\author{B.~A.~Brown$^{1,4}$}
\author{M.~Cortesi$^{1}$}
\author{C.~Fry$^{1,4}$}
\author{B.~E.~Glassman$^{1,4}$}
\author{J.~Heideman$^{7}$}
\author{M.~Janasik$^{1,4}$}
\author{A.~Kruskie$^{1,4}$}
\author{A.~Magilligan$^{1,4}$}
\author{M.~Roosa$^{1,4}$}
\author{J.~Stomps$^{1,4}$}
\author{J.~Surbrook$^{1,4}$}
\author{P.~Tiwari$^{1,4}$}


\affiliation{\footnotesize
$^1$National Superconducting Cyclotron Laboratory, Michigan State University, East Lansing, Michigan 48824, USA\\
$^2$School of Physics and Astronomy, Shanghai Jiao Tong University, Shanghai 200240, China\\
$^3$Racah Institute of Physics, Hebrew University, Jerusalem 91904, Israel\\
$^4$Department of Physics and Astronomy, Michigan State University, East Lansing, Michigan 48824, USA\\
$^5$Canadian Nuclear Laboratories, Chalk River, Ontario K0J 1J0, Canada\\
$^6$IRFU, CEA, Universit\'{e} Paris-Saclay, F-91191, Gif-sur-Yvette, France\\
$^7$Department of Physics and Astronomy, University of Tennessee, Knoxville, Tennessee 37996, USA\\
}

\begin{abstract}
\begin{description}
\item[Background] $\beta$-decay spectroscopy provides valuable information on exotic nuclei and a stringent test for nuclear theories beyond the stability line.
\item[Purpose] To search for new $\beta$-delayed protons and $\gamma$ rays of $^{25}$Si to investigate the properties of $^{25}$Al excited states.
\item[Method] $^{25}$Si $\beta$ decays were measured by using the Gaseous Detector with Germanium Tagging system at the National Superconducting Cyclotron Laboratory. The protons and $\gamma$ rays emitted in the decay were detected simultaneously. A Monte Carlo method was used to model the Doppler broadening of $^{24}$Mg $\gamma$-ray lines caused by nuclear recoil from proton emission. Shell-model calculations using two newly developed universal \textit{sd}-shell Hamiltonians, USDC and USDI, were performed.
\item[Results] The most precise $^{25}$Si half-life to date has been determined. A new proton branch at 724(4)~keV and new proton-$\gamma$-ray coincidences have been identified. Three $^{24}$Mg $\gamma$-ray lines and eight $^{25}$Al $\gamma$-ray lines are observed for the first time in $^{25}$Si decay. The first measurement of the $^{25}$Si $\beta$-delayed $\gamma$ ray intensities through the $^{25}$Al unbound states is reported. All the bound states of $^{25}$Al are observed to be populated in the $\beta$ decay of $^{25}$Si. Several inconsistencies between the previous measurements have been resolved, and new information on the $^{25}$Al level scheme is provided. An enhanced decay scheme has been constructed and compared to the mirror decay of $^{25}$Na and the shell-model calculations.
\item[Conclusions] The measured excitation energies, $\gamma$-ray and proton branchings, log~$ft$ values, and Gamow-Teller transition strengths for the states of $^{25}$Al populated in the $\beta$ decay of $^{25}$Si are in good agreement with the shell-model calculations, offering gratifyingly consistent insights into the fine nuclear structure of $^{25}$Al.
\end{description}
\end{abstract}

\maketitle

\section{Introduction}
The investigation of exotic nuclei lying far from the stability line has been one of the attractive topics of nuclear physics during the past few decades~\cite{Pfutzner_RMP2012}. $\beta$-decay studies have proved to be a powerful tool to obtain a variety of spectroscopic information on nuclei far from stability that are difficult to obtain otherwise~\cite{Blank_PPNP2008,Borge_PS2013}, which provides an excellent and stringent test of nuclear structure theories and fundamental symmetries~\cite{Hardy_PRC2015} and also deepens our understanding of the astrophysical rapid proton capture process~\cite{Wallace_APJS1981}, rapid neutron capture process~\cite{Kajino_PPNP2019}, and $p$ process~\cite{Arnould_PR2003}.

Nuclei near the proton drip line with large $Q$ values for $\beta^+$ decay and low proton separation energies often decay by $\beta$-delayed proton emission ($\beta p$). Since the discovery of the first $\beta p$ emitter $^{25}$Si in 1963, a total of 196 $\beta p$ emitters (including isomers) have been identified ranging from C ($Z = 6$) to Lu ($Z = 71$)~\cite{Batchelder_ADNDT2020}. The $\beta$ decay of $^{25}$Si has been one of the most studied cases~\cite{Barton_CJP1963,McPherson_CJP1965,Hardy_CJP1965,Reeder_PR1966,Sextro_Thesis1973,Zhou_PRC1985,Garcia_PRC1990,Hatori_NPA1992,Robertson_PRC1993,Thomas_EPJA2004}. All the $\beta$-decay measurements of $^{25}$Si were focused on the proton spectrum, whereas the $\gamma$-ray spectrum has not been measured with high statistics. Construction of the decay scheme based solely on proton spectra could lead to inaccurate assignments. Thomas~\textit{et al}.~\cite{Thomas_EPJA2004} reported the most comprehensive measurement but with very limited $\gamma$-ray information. They may have missed some of the low-intensity and high-energy $\gamma$ rays due to low statistics and low efficiency. The existing information on $^{25}$Si decay properties is still incomplete and therefore motivates new experiments to search for new $\beta$-delayed particles and $\gamma$ rays. Detecting protons and $\gamma$ rays in $^{25}$Si $\beta$ decay and the coincidence between them allows one to reliably construct the decay scheme. $^{25}$Si $\beta$-decay spectroscopy provides a sensitive and selective means to probe the properties of $^{25}$Al excited states as well as a good verification of the information on the structure of $^{25}$Al previously collected by other experimental approaches.

It should be noted that most of the information on the $\beta$-delayed proton decay of $^{25}$Si was obtained with silicon implantation detectors. A major problem for this method is the strong $\beta$-summing effect caused by energy deposited by $\beta$ particles~\cite{Blank_PPNP2008}. Robertson~\textit{et al}. employed a gas-silicon detector telescope to detect $^{25}$Si $\beta$-delayed protons for the first time. Despite the small solid angle coverage and the existence of dead layers for incident particles, they were able to identify several new low-energy proton peaks~\cite{Robertson_PRC1993}. Hence, the development of complementary experimental tools for the clean detection of low-energy $\beta$-delayed proton branches is particularly valuable.

In this paper, the emitted particles and $\gamma$ rays in the $\beta$ decay of $^{25}$Si were measured simultaneously with high efficiency and high energy resolution. Combining all available experimental information yields an improved decay scheme of $^{25}$Si, which is compared to theoretical calculations and to the $\beta^-$ decay of the mirror nucleus, $^{25}$Na. A comparison between the mirror Gamow-Teller decays also provides an opportunity to investigate isospin asymmetry. A nonzero mirror asymmetry parameter implies abnormal nuclear structure, such as halo structure in the initial and/or final state. In view of the asymmetries reported in the nearby $sd$-shell nuclei $^{20}$Mg$-^{20}$O~\cite{Piechaczek_NPA1995,Lund_EPJA2016,Sun_PRC2017,Alburger_PRC1987}, $^{22}$Si$-^{22}$O~\cite{Lee_PRL2020,Weissman_JPG2005}, $^{24}$Si$-^{24}$Al~\cite{Ichikawa_PRC2014,Ichikawa_JPConf2011,McDonald_PR1969} $^{26}$P$-^{26}$Na~\cite{Perez-Loureiro_PRC2016,Grinyer_PRC2005}, $^{27}$S$-^{27}$Na~\cite{Janiak_PRC2017,Sun_PRC2019,Detraz_PRC1979,Guillemaud-Mueller_NPA1984}, it is desirable to extend this test to $^{25}$Si and its mirror partner nucleus $^{25}$Na~\cite{Jones_PRC1970,Alburger_PRC1971,Alburger_NPA1982}.

\section{Experimental techniques}
The experiment was conducted at the National Superconducting Cyclotron Laboratory (NSCL) in May 2018. The experimental procedure has been detailed in Ref.~\cite{Friedman_NIMA2019} and is briefly repeated here for completeness. A $^{36}$Ar$^{18+}$ primary beam was accelerated by the K500 and K1200 Coupled Cyclotron Facility to 150~MeV/nucleon at a beam current of $\sim$75~$p$nA. The secondary $^{25}$Si beam was produced via the projectile fragmentation of the $^{36}$Ar beam impinging on a 1363~$\mathrm{mg/cm^2}$ thick $^9$Be target and purified using the A1900 fragment separator~\cite{Morrissey_NIMB2003}. The Gaseous Detector with Germanium Tagging (GADGET)~\cite{Friedman_NIMA2019}, composed of the Proton Detector and the Segmented Germanium Array (SeGA)~\cite{Mueller_NIMA2001}, has been built and successfully commissioned to measure the decays for the nuclei near the proton-drip line. In the current experiment, a total of $3\times10^{7}$ $^{25}$Si ions were implanted into the gaseous Proton Detector with an average beam rate of approximately 1800~particles per second. The Proton Detector was filled with P10 gas mixture at a pressure of 780~Torr, which is ideally suited for low-energy proton detection because the background contributed by $\beta$ particles was mitigated. The charged-particle measurement was carried out under a pulsed-beam mode, i.e., the beam ions were delivered for 500~ms, then the decays were detected during the 500-ms beam-off period. The Proton Detector was mounted at the center of SeGA, which consists of 16 high-purity germanium detectors arranged into two rings surrounding the Proton Detector. These two rings of eight detectors will be referred to as ``upstream'' and ``downstream''. The detection for the $\gamma$ rays emitted from decays was done over both the beam on/off periods. The preamplifier signals from the Proton Detector and SeGA were read into Pixie-16 cards (16-Channel 250-MHz PXI Digital Processor) and processed by the NSCL digital data acquisition system~\cite{Prokop_NIMA2014}.
\section{Analysis}
\subsection{$\gamma$-ray energy and efficiency calibration}
To create a cumulative $\gamma$-ray energy spectrum, the spectrum of each SeGA detector was linearly gain-matched run by run using room background lines at $1460.820\pm0.005$ and $2614.511\pm0.010$~keV from the $\beta$ decays of $^{40}$K~\cite{Chen_NDS2017} and $^{208}$Tl~\cite{Martin_NDS2007}, respectively.
An exponentially modified Gaussian (EMG) function of the form
\begin{equation}
\begin{split}
f(x;N,\mu,\sigma,\tau)=\frac{N}{2\tau}\mathrm{exp}\left[\frac{1}{2}\left(\frac{\sigma}{\tau}\right)^2+\frac{x-\mu}{\tau}\right]\\
\times\mathrm{erfc}\left[\frac{1}{\sqrt{2}}\left(\frac{\sigma}{\tau}+\frac{x-\mu}{\sigma}\right)\right],
\end{split}
\end{equation}
was used to fit each $\beta$-delayed $\gamma$-ray line in the spectrum. The EMG is characterized by an exponential decay constant $\tau$, width $\sigma$, mean $\mu$, energy $x$, and area below the curve $N$. Also, a linear function is added to this formula to model the local background. Four $^{25}$Si $\beta$-delayed $\gamma$-ray lines with known energies and the corresponding absolute intensities shown in brackets--451.7(5)~keV [18.4(42)\%], 493.3(7)~keV [15.3(34)\%], 944.9(5)~keV [10.4(23)\%], and 1612.4(5)~keV [14.7(32)\%]~\cite{Firestone_NDS2009_25,Piiparinen_ZP1972,Thomas_EPJA2004}--were observed with high statistics and used as energy calibration standards. The maximum values from the fits of these $\gamma$-ray lines were plotted against the standard energies to provide an internal energy calibration of each SeGA detector. In this paper, all the $\gamma$-ray energies are reported in the laboratory frame, and all the excitation energies are reported in the center-of-mass frame with recoil corrections applied. One of the 16 SeGA detectors malfunctioned during the experiment and three of the others displayed relatively poor resolutions, so these four detectors are excluded from the subsequent analysis. A cumulative spectrum incorporating the other 12 SeGA detectors was generated for analysis. The characteristic resolution for the cumulative SeGA spectrum is 0.19\% full width at half maximum at 1612~keV.

To reduce the systematic uncertainty associated with extrapolation, further energy calibration was applied by including four $^{25}$Si($\beta p\gamma$)$^{24}$Mg lines known with very good precision at 1368.626(5), 2754.007(11), 2869.50(6), and 4237.96(6)~keV as standards~\cite{Firestone_NDS2007_24}. These $\gamma$ rays are emitted from the recoiling $^{24}$Mg after the $\beta$-delayed proton emission of $^{25}$Si. Therefore, the $\gamma$-ray line shape is Doppler broadened, and the regular EMG function is not suited to fit the peak. To accurately extract information from each peak, we applied the Doppler broadening line shape analysis. The detailed procedure will be described in Sec.~\ref{Doppler broadening analysis}.

\subsection{Proton energy and efficiency calibration}
As detailed in Ref.~\cite{Friedman_NIMA2019}, the anode plane of the Proton Detector is segmented into 13 readout pads, labeled A$-$M. The $\beta$-delayed proton spectrum is usually produced by event-level summing of the five central pads (A$-$E) and the eight surrounding pads (F$-$M) are usually used to veto the high-energy protons that escape the active volume and deposit only part of their energy in the active volume. In the current experiment, four veto pads (F, G, L, M) were not instrumented, so the resulting background caused by the escaping high-energy protons hindered the identification of low-energy protons. Instead, we could obtain the proton spectrum measured by three central pads (A+C+D) and used the other six neighboring pads (B, E, H, I, J, K) as veto triggers. The strong $\beta$-delayed proton peaks at 402, 1268, and 1924~keV were used for the energy calibration of the Proton Detector. We took a weighted average of the literature proton energies~\cite{Sextro_Thesis1973,Hatori_NPA1992,Robertson_PRC1993,Thomas_EPJA2004} as calibration standards. The proton information in our paper is incomplete compared with other literature as our Proton Detector is not sensitive to protons above 2.4~MeV. Besides, the proton-detection efficiency simulated for full utilization of all 13 readout pads~\cite{Friedman_NIMA2019} cannot be used in this case. It is simpler and more accurate to determine the intensities for each proton branch by normalizing the literature relative intensities to the $^{25}$Si($\beta p\gamma$)$^{24}$Mg intensities measured in this paper (Sec.~\ref{25Sibpg24Mg}).

\subsection{\label{Normalization}Normalization}
In Ref.~\cite{Friedman_NIMA2019}, we investigated the longitudinal beam distribution via the proton drift time distribution, and the beam in the radial direction was estimated as a Gaussian beam with the transverse distribution determined based on the distribution of proton counts in different pads of the Proton Detector. The investigation showed that the $^{25}$Si beam ions were mainly contained in the active volume of the Proton Detector. We modeled the Brownian motion of the $^{25}$Si atoms using a Monte Carlo simulation. The diffusion of the $^{25}$Si atoms is estimated to be less than 1~cm within four lifetimes, and there was very little drift of $^{25}$Si ions to the cathode of the Proton Detector. The $\beta$-delayed $\gamma$ rays and protons from subsequent decays were detected by the SeGA detectors and the Proton Detector, respectively. The geometry of our experimental setup and the beam spatial distribution were used as inputs for a \textsc{geant}4~\cite{Agostinelli_NIMA2003,Allison_NIMA2016} Monte Carlo simulation to determine a $\gamma$-ray photopeak efficiency curve for the SeGA detectors. We then verified the simulated efficiency curve by using a $^{152}$Eu calibration source~\cite{Martin_NDS2013} between 245 and 1408~keV and $^{23}$Al data~\cite{Zhai_Thesis2007} up to 7801~keV. The source data were taken with the $^{152}$Eu source placed at the center of SeGA before the Proton Detector was installed and the $^{23}$Al($\beta\gamma$)$^{23}$Mg was measured using the same detection setup in a subsequent experiment in the same campaign~\cite{Friedman_PRC2020}. Although the $^{152}$Eu source was absolutely calibrated, our procedure for determining the absolute intensities of the $\gamma$ rays only requires relative efficiencies. We extracted the $\gamma$-ray efficiency from both data and simulation, and the simulated efficiency is matched with the measured efficiency when scaled by a constant factor on the order of unity. The uncertainties associated with the scaling factors are determined to be 0.7\% for $\gamma$-ray energies $<$1.4~MeV based on the $^{152}$Eu source data and 4.2\% for $\gamma$-ray energies $>$1.4~MeV based on the $^{23}$Al data, which give a measure of the uncertainty on the relative efficiency. We then add a flat 2\% uncertainty in the efficiencies at all energies to account for the $\gamma\gamma$ summing effect~\cite{Glassman_PRC2019}. Ultimately, we adopt a conservative 3\% uncertainty envelope for $\gamma$-ray energies $<$1.4~MeV and a 5\% uncertainty envelope for $\gamma$-ray energies $>$1.4~MeV. The uncertainties associated with the relative efficiencies were propagated through the calculation of each $\gamma$-ray intensity.

We adopt an $I_{\mathrm{gs}}=21.5(12)\%$ $\beta$ feeding for the $^{25}$Al ground state based on our shell-model calculated $I_{\mathrm{gs}}=20.9$ and 22.2\%. The difference between the two theoretical $I_{\mathrm{gs}}$ represents the uncertainty coming from the Hamiltonian (Sec.~\ref{Shell-model calculations}). We can perform the normalization by requiring the intensity of all decay paths sum to 100\%:

\begin{equation}
\begin{split}
I_{\beta p}+I_{\beta\gamma}+I_{\mathrm{gs}}=100\%,\\
 \frac{I_{\beta p\gamma}}{I_{\beta p}}=59.0(5)\%,
\end{split}
\end{equation}

where $I_{\beta p}$ is the total intensity of all $^{25}$Si($\beta p$)$^{24}$Mg transitions, and $I_{\beta\gamma}$ is the total intensity of all $^{25}$Si($\beta\gamma$)$^{25}$Al transitions. The intensity of the $^{25}$Si($\beta p\gamma$)$^{24}$Mg ($I_{\beta p\gamma}$) accounts for 59.0(5)\% of the total $^{25}$Si($\beta p$)$^{24}$Mg intensity based on the previous $\beta$-delayed proton measurements~\cite{Hatori_NPA1992,Robertson_PRC1993,Thomas_EPJA2004}. The remainder of our normalization procedure is entirely based on the $\gamma$-ray intensities. Using the simulated relative efficiency and the number of counts in each peak extracted from the EMG fits yields the intensity for each $\gamma$ ray. Multiple $\gamma$ rays in one cascade are treated as one transition. Thus, we determine the total intensities of the $\beta$-delayed proton and $\beta$-delayed $\gamma$ decays to be $I_{\beta\gamma}=40.1(14)\%$ and $I_{\beta p}=38.3(15)\%$, respectively. These values can be converted into the $\beta$ feedings to all the unbound $^{25}$Al states $I_{\mathrm{unb}}=39.3(15)\%$ and all the bound $^{25}$Al states $I_{\mathrm{bnd}}=60.7(18)\%$ when taking into account the weak $\gamma$-ray intensities originating from the unbound states (Sec.~\ref{25Sibg25Al}). Our values may be compared with the previous literature values of $I_{\mathrm{unb}}=40.5(14)\%$~\cite{Hatori_NPA1992}, 37.7(15)\%~\cite{Robertson_PRC1993}, and 35.2(12)\%~\cite{Thomas_EPJA2004} and $I_{\mathrm{bnd}}=58.7(13)\%$~\cite{Hatori_NPA1992}, 61.9(26)\%~\cite{Robertson_PRC1993}, and 66(9)\%~\cite{Thomas_EPJA2004}.

\subsection{\label{Doppler broadening analysis}Doppler broadening analysis}
When a proton is emitted from a nucleus, the daughter nucleus will recoil with equal and opposite momentum as the ejected proton due to the conservation of momentum. If a $\gamma$ ray is emitted while the nucleus is still recoiling, it will be Doppler shifted in the laboratory frame. For an ensemble of such events, the resulting $\gamma$-ray line shape in the measured energy spectrum will be Doppler broadened. In this experiment, we observed four $\gamma$-ray lines emitted from the $^{24}$Mg recoiling in the gas after the $\beta$-delayed proton emission of $^{25}$Si. Detailed Monte Carlo simulations have been developed to model the Doppler broadening. The results are then compared to the actual $\gamma$-ray data~\cite{Fynbo_NIMB2003,Fynbo_NPA2004,Sarazin_PRC2004,Mattoon_PRC2009,Schwartz_PRC2015,Glassman_PRC2019}. The simulation takes into account the energy and relative intensity of each proton branch populating the $^{24}$Mg excited state, the energy of the $\gamma$ ray deexciting the $^{24}$Mg excited state, the lifetime of the $^{24}$Mg excited state, the stopping power of the implantation material (780-Torr P10 gas), and the response function of each SeGA detector.

Robertson~\textit{et al}.~\cite{Robertson_PRC1993} and Thomas~\textit{et al}.~\cite{Thomas_EPJA2004} reported the most comprehensive $^{25}$Si($\beta p$)$^{24}$Mg assignments and they are generally in agreement. Hence, we adopted their proton energies and proton feeding intensities in the simulation. The stopping power of the recoiling $^{24}$Mg in P10 gas is estimated as a function of energy using the code \textsc{srim}, which is expected to be accurate to within 10\%~\cite{Ziegler_NIMB2010}. The lifetimes for the three low-lying $^{24}$Mg excited states at 1368, 4123, and 4238~keV have been precisely measured to be 1.92(9)~ps, 31.7(3)~fs, and 59.2(6)~fs, respectively~\cite{Firestone_NDS2007_24}. An isotropic distribution of $\gamma$ rays with respect to the proton distribution is assumed in each simulation. Another input of the simulation is the intrinsic response function for each of the SeGA detectors. By fitting unbroadened $\beta$-delayed $\gamma$-ray peaks with the EMG function Eq.~(1) at energies of 451.7(5), 493.3(7), 944.9(5), and 1612.4(5)~keV [$^{25}$Si($\beta\gamma$)$^{25}$Al]~\cite{Firestone_NDS2009_25} and 450.70(15), 1599(2), 2908(3), and 7801(2)~keV [$^{23}$Al($\beta\gamma$)$^{23}$Mg]~\cite{Firestone_NDS2007_23}, the parameters $\tau$ and $\sigma$ were characterized as a function of energy for each SeGA detector. Every detector has a different contribution to the total number of counts in the peak depending on its detection efficiency, and the simulation accounts for this by normalizing the number of counts simulated for each detector.


A linear function is adopted to model the local background and added to each simulated peak when compared to the actual data. Then, the simulation-data comparison can be done using the classical $\chi^2$-minimization method. Because of the relatively low statistics collected in the present experiment compared to the $^{20}$Mg($\beta p\gamma$)$^{19}$Ne experiment~\cite{Glassman_PRC2019}, the construction of a simulated peak shape follows the same method of Ref.~\cite{Glassman_PRC2019} with one major change. Although least-squares based $\chi^2$ statistics (e.g., Neyman's $\chi^2$ or Pearson's $\chi^2$) are widely used for this type of analysis, they do not always give reliable results for low-statistics data. An alternative method better suited for low-statistics analysis is to derive a $\chi^2$ statistic from a Poisson maximum likelihood function~\cite{Baker_NIM1984,Chester_PRC2017}. The ``likelihood $\chi^2$" is defined in the equation

\begin{equation}
\chi^2=-2\mathrm{ln}\lambda=2\sum_{i=1}^Ny_i-n_i+n_i\mathrm{ln}\frac{n_i}{y_i},
\end{equation}

where $\lambda$ is the likelihood ratio, $N$ is the number of bins, $n_i$ is the number of counts in the $i$th bin of the measured spectrum, and $y_i$ is the number of counts in the $i$th bin predicted by the simulation. The minimization of $\chi^2$ is equivalent to the maximization of $\lambda$. The binned maximum likelihood method is known to miss the information with feature size smaller than the bin size of the spectrum. It is therefore necessary to use a fine binning for line-shape analysis. We used a 0.1-keV bin size for the $\gamma$-ray spectrum. The application of this likelihood ratio $\chi^2$ method to the $^{25}$Si($\beta p\gamma$)$^{24}$Mg line-shape analysis is discussed in Sec.~\ref{25Sibpg24Mg}.

\section{Results and Discussion}
\subsection{Half-life of $^{25}$Si}
\begin{figure}
\begin{center}
\includegraphics[width=8.6cm]{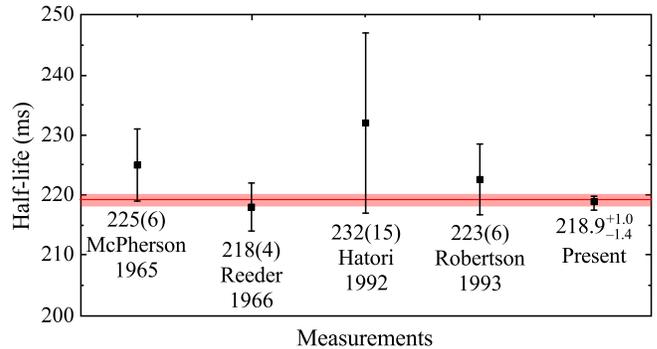}
\caption{\label{FT}Half-life of $^{25}$Si measured in the present paper compared with the values previously measured by McPherson and Hardy~\cite{McPherson_CJP1965}, Reeder \textit{et al}.~\cite{Reeder_PR1966}, Hatori \textit{et al}.~\cite{Hatori_NPA1992}, and Robertson \textit{et al}.~\cite{Robertson_PRC1993}. The weighed average of all published values is indicated by the solid red band.}
\end{center}
\end{figure}

The $^{25}$Si half-life has been previously measured to be 225(6)~ms~\cite{McPherson_CJP1965}, 218(4)~ms~\cite{Reeder_PR1966}, 232(15)~ms~\cite{Hatori_NPA1992}, and 222.6(59)~ms~\cite{Robertson_PRC1993}. A weighted average of all previously published values gives $t_{1/2}=221.1(28)$~ms. In Ref.~\cite{Friedman_NIMA2019}, we have shown the decay curve of $^{25}$Si by using the count rate of the 402-keV proton as a function of time during the decay period of the implant-decay cycle. Here, we further investigated the systematics to provide a half-life measurement. The $^{25}$Si half-life is extracted by fitting the count rate of all the protons within 350$-$2400~keV recorded by the five central pads as a function of time elapsed since the beginning of each implant-decay cycle. The decay in the count rate is enhanced by diffusion of $^{25}$Si out of the active volume. This effect is modeled by a Monte Carlo simulation of the Brownian motion of the $^{25}$Si atoms. The effect of $^{25}$Si losses due to diffusion out of the central pads is parametrized by a fourth degree polynomial $P_{4}(t-t_0)$ where $t$ is the clock time within the cycle and $t_0$ is the beginning of the decay period of the cycle. The data are fit using the function

\begin{equation}
  f(t;N,t_{1/2},t_0,B) = Ne^{-\frac{\ln(2)(t-t_0)}{t_{1/2}}}P_4(t-t_0)+B,
\end{equation}

where $N$ is the initial count rate of protons at the beginning of the decay period of the cycle. We measured the background during the interval between each run and estimated the background level $B$ to be $0.82(2)$ count/s. The systematic effect associated with the uncertainty on $t_0$ is estimated to be 0.05~ms, and the systematic effect associated with the fit range is estimated to be $\pm0.9$~ms by varying $t_0$ and the fit range within reasonable values. The diffusion is estimated to decrease the decay lifetime by 1.8~ms. However, this assumes that the $^{25}$Si is in the atomic form. In reality, it is plausible that Si atoms bond with hydrogen and carbon atoms that exist in the P10 mixture. As a result, the diffusion is expected to decrease in a nontrivial manner. We then estimate other systematic effects due to the diffusion by varying the initial $^{25}$Si beam distribution in the volume and the gas pressure of the simulation. The total uncertainty associated with diffusion is determined to be $^{+0.3}_{-0.9}$~ms. Other effects, such as the trigger threshold, the time window for the trigger, and the contribution of the beam contaminants, are found to be negligible. The effects contributing to the uncertainty are summarized in Table~\ref{Thalflife}. The final result is determined to be $t_{1/2}=218.9\pm0.5(\mathrm{stat})^{+0.9}_{-1.3}(\mathrm{syst})$~ms, where the uncertainties are statistical and systematic, respectively. This value can be written as $t_{1/2}=218.9^{+1.0}_{-1.4}$~ms with the statistical and systematic uncertainties added in quadrature. As can be seen from Fig.~\ref{FT}, our result is consistent with, and more precise than, all the literature values~\cite{McPherson_CJP1965,Reeder_PR1966,Hatori_NPA1992,Robertson_PRC1993}. We have reevaluated the half-life to be $t_{1/2}=219.2^{+0.9}_{-1.2}$~ms by taking a weighted average of all published values.

\begin{table}
\caption{\label{Thalflife} Uncertainty budget for the measured half-life of $^{25}$Si.}
\begin{center}
\begin{ruledtabular}
\begin{tabular}{cc}
Source of uncertainty & Uncertainty~(ms) \\
\hline
 Statistics                                  & $\pm0.5$ \\
 Diffusion of $^{25}$Si atoms   & $^{+0.3}_{-0.9}$ \\
 Starting time of decay period    & $\pm0.05$ \\
 Fit range                                  & $\pm0.9$ \\
 Trigger threshold                      & negligible \\
 Event window                          & negligible \\
 Contaminants                           & negligible \\
 Total                                       & $^{+1.0}_{-1.4}$ \\
\end{tabular}%
\end{ruledtabular}
\end{center}
\end{table}

\subsection{\label{25Sibpg24Mg}$^{25}$Si($\beta p\gamma$)$^{24}$Mg}

\begin{figure*}
\begin{center}
\includegraphics[width=17cm]{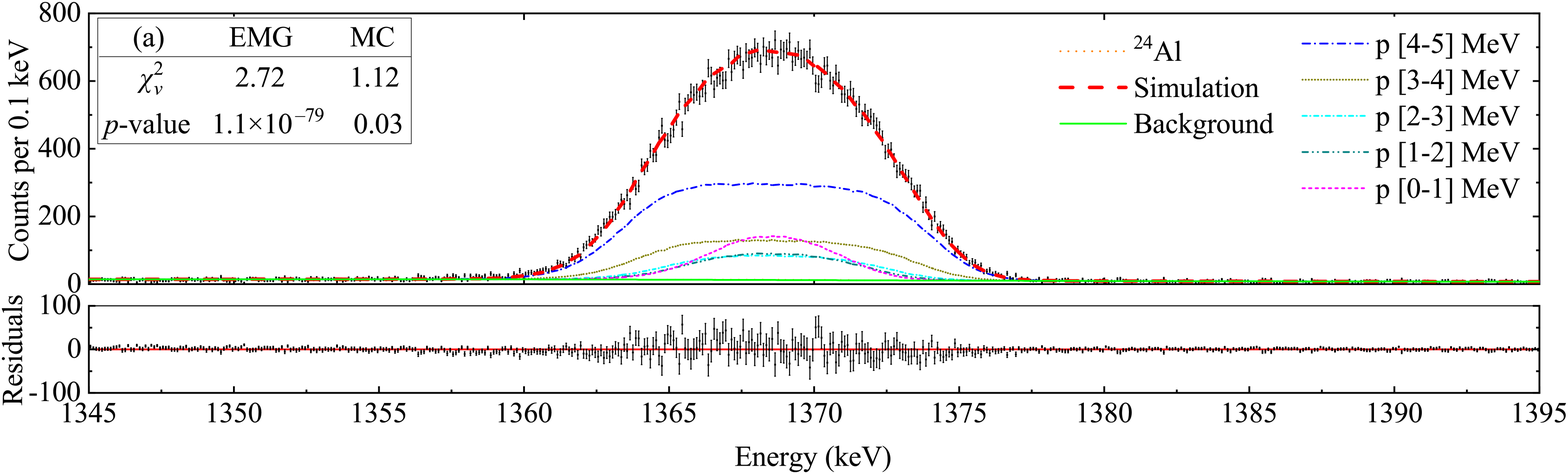}
\includegraphics[width=17cm]{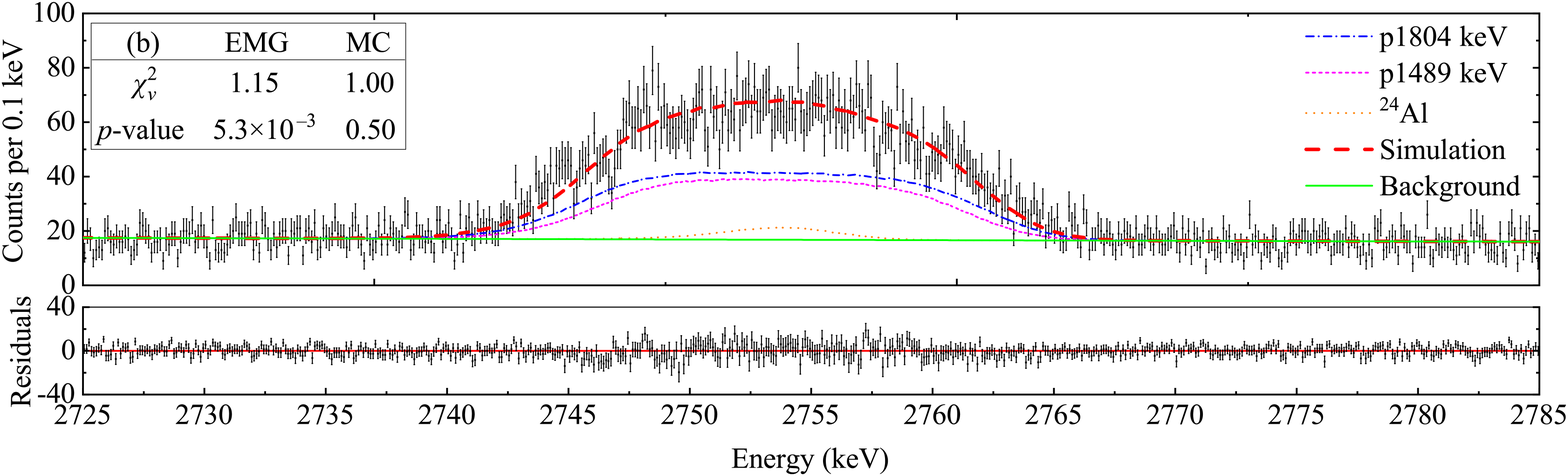}
\includegraphics[width=17cm]{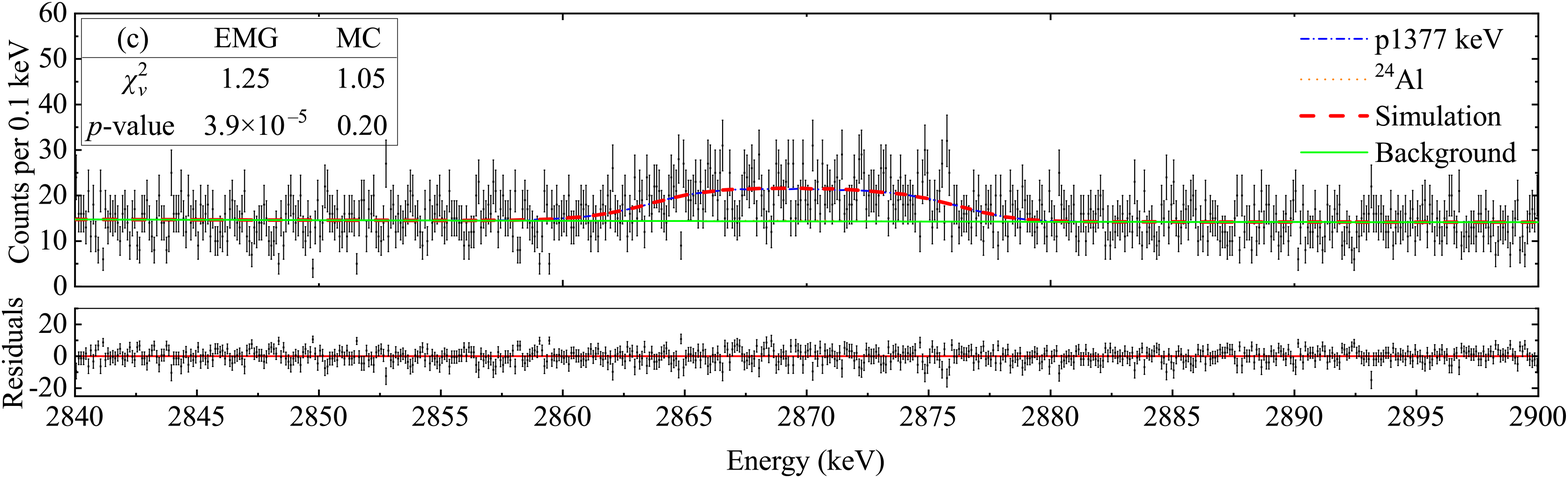}
\includegraphics[width=17cm]{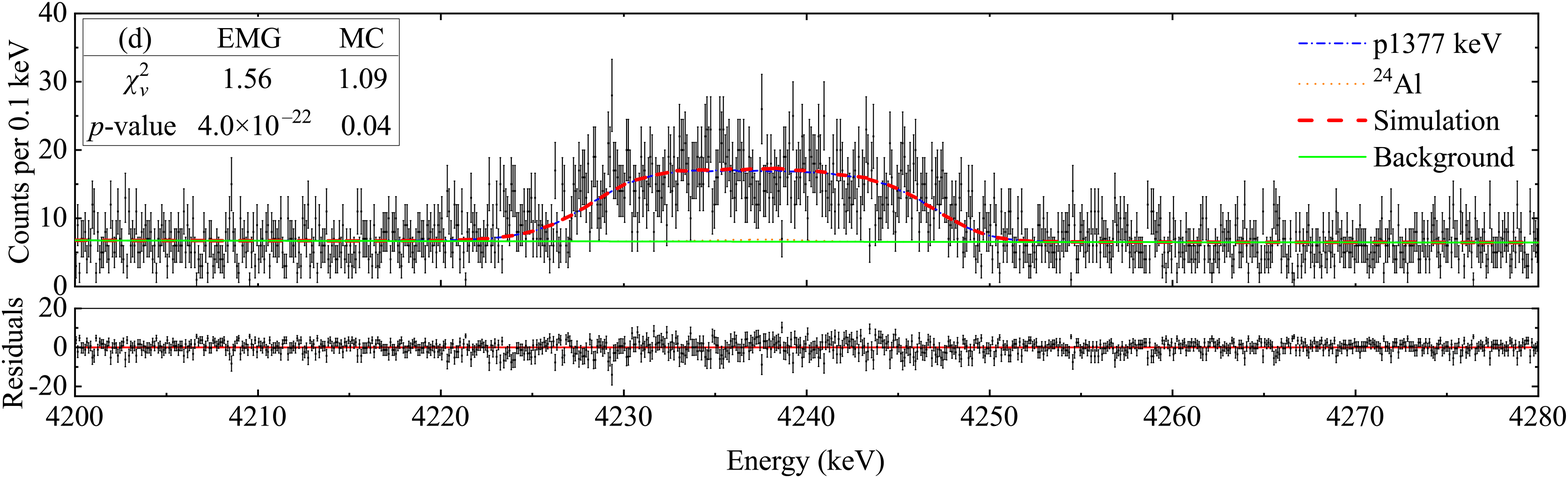}
\caption{\label{Fbpg}$\gamma$-ray spectrum measured by the SeGA detectors magnified at (a) 1369~keV, (b) 2754~keV, (c) 2870~keV, and (d) 4238~keV. We show the raw spectrum for panel (b), (c), and (d) and the Proton-Detector-gated spectrum for panel (a) to suppress the contribution from a room background line near the 1369-keV peak. Four upper panels: The Monte Carlo (MC) simulations of the (a) 1369-keV, (b) 2754-keV, (c) 2870-keV, and (d) 4238-keV $\gamma$-ray peaks are produced by using lifetimes adopted from the data evaluation~\cite{Firestone_NDS2007_24} and the proton energies and relative proton feeding intensities measured by Thomas~\textit{et al}.~\cite{Thomas_EPJA2004}. The black dots represent the data, the solid green lines denote the background model, the dashed red lines denote the simulated line shapes including different contributions of proton feedings. Each proton feeding is represented by a colored line, and in the legend it is labeled with a letter $p$ followed by its center-of-mass energy. The dotted orange lines denote the small unbroadened contribution of the contaminant decay $^{24}$Al($\beta\gamma$)$^{24}$Mg. Four lower panels: The residual plots show the data subtracted from the simulation. Compared with a regular EMG fit of each peak, our Doppler broadening analysis substantially improved the $\chi_\nu^2$ and $p$ values, which are shown in the top-left corner of each panel.}
\end{center}
\end{figure*}

\begin{figure}
\begin{center}
\includegraphics[width=8cm]{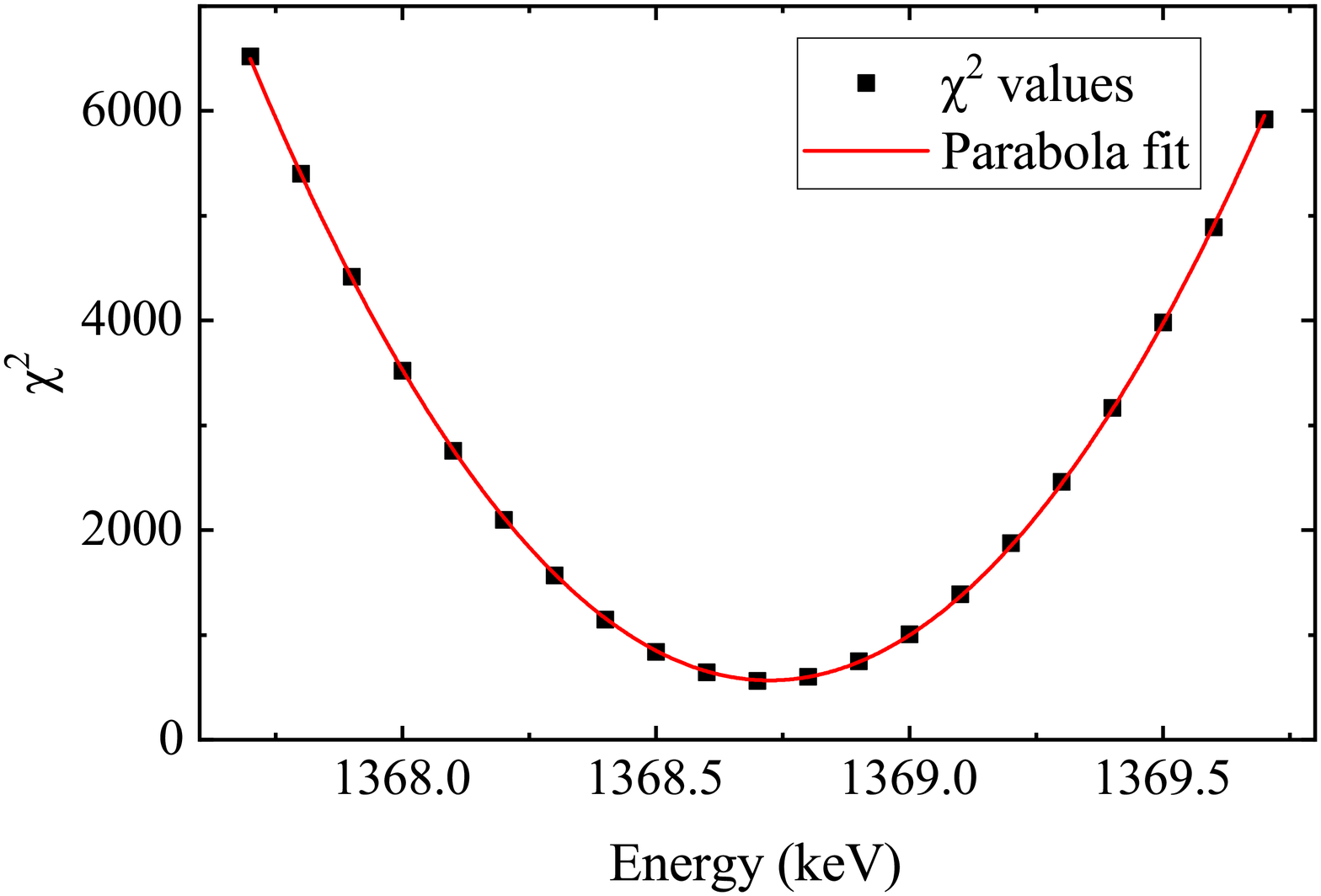}
\caption{\label{Fchi}$\chi^2$ distribution of the 1368.672(5)-keV $\gamma$-ray line as a function of input $\gamma$-ray energy (black squares). A quadratic polynomial fit (solid red line) was used to determine the best-fit energy and uncertainty.}
\end{center}
\end{figure}

A 1369-keV $\gamma$ ray originating from the first excited state of $^{24}$Mg was observed in the previous $^{25}$Si $\beta$-decay measurements~\cite{Garcia_PRC1990,Thomas_EPJA2004}. In this paper, we observed three additional $^{24}$Mg $\gamma$-ray lines following $^{25}$Si $\beta$-delayed proton emissions. Figure~\ref{Fbpg} shows four $\gamma$-ray lines at 1369, 2754, 2870, and 4238~keV, corresponding to the deexcitations from the three lowest-lying $^{24}$Mg excited states populated by $^{25}$Si($\beta p$). The placement of $\gamma$ rays is also verified using $\gamma\gamma$ coincidences. To remove the disturbance from a room background $\gamma$-ray line near the 1369-keV peak, the Proton-Detector-gated $\gamma$-ray spectrum was generated for the 1369-keV peak. This coincidence gate is set by any signal above 15~keV recorded by the Proton Detector, and all the protons emitted from decays essentially have equal probabilities to trigger a 10-$\mu$s backward time window and select the coincident $\gamma$-ray signals. Therefore, the Proton-Detector gate only reduces the number of counts in the 1369-keV $\gamma$-ray line and does not alter its relative proton feedings and the resulting peak shape. Distributions of $\chi^2$ values from the simulated and experimental spectra were constructed for each peak. An example of the $\chi^2$ distribution of the 1369-keV $\gamma$-ray line is shown in Fig.~\ref{Fchi}, where the $\gamma$-ray centroid is considered a free parameter for $\chi^2$ minimization. The best-fit peak centroid and integral as well as their statistical uncertainties ($\chi_\mathrm{min}^2+1$) were taken from a quadratic polynomial fit of the $\chi^2$ distribution. We obtained the reduced $\chi^2$ value ($\chi_\nu^2$) by dividing the $\chi^2$ value by the number of degrees of freedom. Each statistical uncertainty is then inflated by the square root of the $\chi_\nu^2$ value for the corresponding fit. We are able to achieve a minimum in the $\chi_\nu^2$ distribution close to 1 for all four $^{24}$Mg $\gamma$-ray lines, using the proton energies and the relative proton feeding intensities measured by Thomas~\textit{et al}.~\cite{Thomas_EPJA2004}. The resultant best fits from the $\chi^2$ minimization are shown in Fig.~\ref{Fbpg}. Replacing the proton energies and the relative proton feeding intensities with the values measured by Robertson~\textit{et al}.~\cite{Robertson_PRC1993} in our simulation yields very similar $\chi^2$ values. Our Doppler broadening analysis does not have sufficient sensitivity to distinguish discrepancies in the intensities of weak proton branches in this relatively low-statistics case. Nevertheless, their proton inputs both fit the $\gamma$-ray data equally well, indicating that both of the previous measurements placed the majority of the proton intensity in the decay scheme correctly.

The $\gamma$-ray intensities per $^{25}$Si $\beta$ decay ($I_{\beta p\gamma}$) are derived from the integral of each fit corrected for the SeGA efficiency and normalized to the aforementioned total $\gamma$-ray intensity. The lifetime and the fit parameters $\tau$ and $\sigma$ are varied by their one standard deviation uncertainty and the stopping power is varied up and down by 50\% to investigate the systematic uncertainty. The uncertainties associated with the aforementioned simulated efficiency, the stopping power of the P10 gas, the lifetime of $^{24}$Mg states, the proton feedings, the parameters $\tau$ and $\sigma$, and the deviation caused by adopting proton energies and intensities from different literature~\cite{Robertson_PRC1993,Thomas_EPJA2004} were added in quadrature to obtain the total systematic uncertainty. Adding the systematic uncertainties with the statistical uncertainty in quadrature yields the total uncertainty for each $\gamma$-ray intensity.


The $^{24}$Al($\beta\gamma$)$^{24}$Mg decay from the beam contaminant $^{24}$Al might yield a small portion of counts in the $^{25}$Si($\beta p\gamma$)$^{24}$Mg peaks as they both produce $^{24}$Mg excited states. The $^{24}$Al($\beta\gamma$)$^{24}$Mg lines are unbroadened and should also be included in the Doppler broadening simulation. A 7070-keV $\gamma$-ray peak is identified in the spectrum, and it can only be produced by $^{24}$Al($\beta\gamma$)$^{24}$Mg. Based on the number of counts and intensity of the 7070-keV $\gamma$ ray~\cite{Warburton_PRC1981}, we estimate that the beam contaminant $^{24}$Al comprised 0.13(4)\% of the implanted beam ions. Thus, the number of counts in the 1369-, 2754-, 2870-, and 4238-keV $\gamma$-ray peaks contributed by $^{24}$Al($\beta\gamma$)$^{24}$Mg are quantified based on the SeGA efficiency at these energies and their known intensities relative to the 7070-keV $\gamma$-ray peak~\cite{Warburton_PRC1981,Firestone_NDS2007_24}.

For the 4238-keV $^{24}$Mg state, a further correction is required. Since the 4238-keV state has two $\gamma$ decay paths to the ground state, the 2870-1369 cascade that does not directly decay to the ground state will yield a small portion of counts in the 4238-keV peak due to summing in a single SeGA detector. The number of counts in the 4238-keV peak due to the summing effect is calculated from the number of counts in the 2870-keV peak and the SeGA efficiency for a 1369-keV $\gamma$ ray. The loss of photopeak counts for the 1369- and 2870-keV $\gamma$ rays due to the summing effect is corrected likewise. After correcting the contaminant counts and the summing counts for the $^{25}$Si($\beta p\gamma$)$^{24}$Mg peak integral, we determine the final $^{25}$Si($\beta p\gamma$)$^{24}$Mg intensities and $\gamma$-ray branching ratios (see Table~\ref{T24Mg}).

The two $\gamma$ rays emitted from the 4238.24(3)-keV $^{24}$Mg state at 4237.96(6) and 2869.50(6)~keV are measured to have branching ratios of 75(3) and 25(3)\%, respectively. The branching ratios are in agreement with the evaluated values of 78.2(10) and 21.8(10)\%~\cite{Endt_NPA1990}, which took a weighed average of the results in Refs.~\cite{Warburton_PRC1981,Endt_NPA1978} with inflated uncertainty. We obtain the $\beta$-delayed proton feedings to the 1369-, 4123-, 4238-keV $^{24}$Mg states per $^{25}$Si decay of $I_{\beta p}=21.0(9)$, 0.94(6), and 0.59(3)\%, respectively, by adding all $\gamma$-ray decays originating from each state and subtracting feeding from higher-lying states. The proton feeding to the $^{24}$Mg ground state accounts for 41.0(5)\% of the total $^{25}$Si($\beta p$)$^{24}$Mg intensity ~\cite{Hatori_NPA1992,Robertson_PRC1993,Thomas_EPJA2004}. Combining this branching ratio and our measured $^{25}$Si($\beta p\gamma$)$^{24}$Mg intensities yields an $I_{\beta p}=15.7(7)\%$ for the $^{24}$Mg ground state. Thomas~\textit{et al}.~\cite{Thomas_EPJA2004} reported $I_{\beta p}=$~14.3(10), 18.7(15), 1.06(20), and 0.41(12)\% and Robertson~\textit{et al}.~\cite{Robertson_PRC1993} reported $I_{\beta p}=$~14.96(5), 20.27(5), 1.105(9), and 0.378(8)\% for the $^{24}$Mg  ground state and excited states at 1369, 4123, and 4238~keV, respectively. The consistency of intensities further confirms the literature interpretation of $^{25}$Si $\beta$-delayed proton branches.

\begin{table}
\caption{\label{T24Mg} $^{25}$Si($\beta p\gamma$)$^{24}$Mg measured in the present paper. The $^{24}$Mg ground state and three lowest-lying excited states are observed to be populated by $^{25}$Si($\beta p$). The well-known $^{24}$Mg excitation energies (column 1) and $\gamma$-ray energies (column 3) are adopted from the data evaluation~\cite{Firestone_NDS2007_24}. Column 2 reports the measured $^{25}$Si($\beta p$)-feeding intensities to each $^{24}$Mg state. Column 4 reports the intensity of each $\gamma$-ray transition per $^{25}$Si decay. Column 5 reports the $\gamma$-ray branching ratios for each $^{24}$Mg excited state. }
\begin{center}
\begin{ruledtabular}
\begin{tabular}{ccccc}
$E_x$ (keV)~\cite{Firestone_NDS2007_24} & $I_{\beta p}$ (\%) & $E_\gamma$ (keV)~\cite{Firestone_NDS2007_24} & $I_{\beta p\gamma}$ (\%) & B.R. (\%) \\
\hline
             0    & 15.7(7) & $-$ & $-$ & $-$  \\
1368.672(5) & 21.0(9) & 1368.626(5) & 22.1(9) & 100 \\
4122.889(12) & 0.94(6) & 2754.007(11) & 0.94(6) & 100 \\
\multirow{2}[0]{*}{4238.24(3)} & \multirow{2}[0]{*}{0.59(3)} & 2869.50(6) & 0.147(14) & 25(3) \\
      &       & 4237.96(6) & 0.44(3) & 75(3) \\
\end{tabular}%
\end{ruledtabular}
\end{center}
\end{table}

\subsection{\label{25Sibg25Al}$^{25}$Si($\beta\gamma$)$^{25}$Al}

\begin{figure*}
\begin{center}
\includegraphics[width=17.2cm]{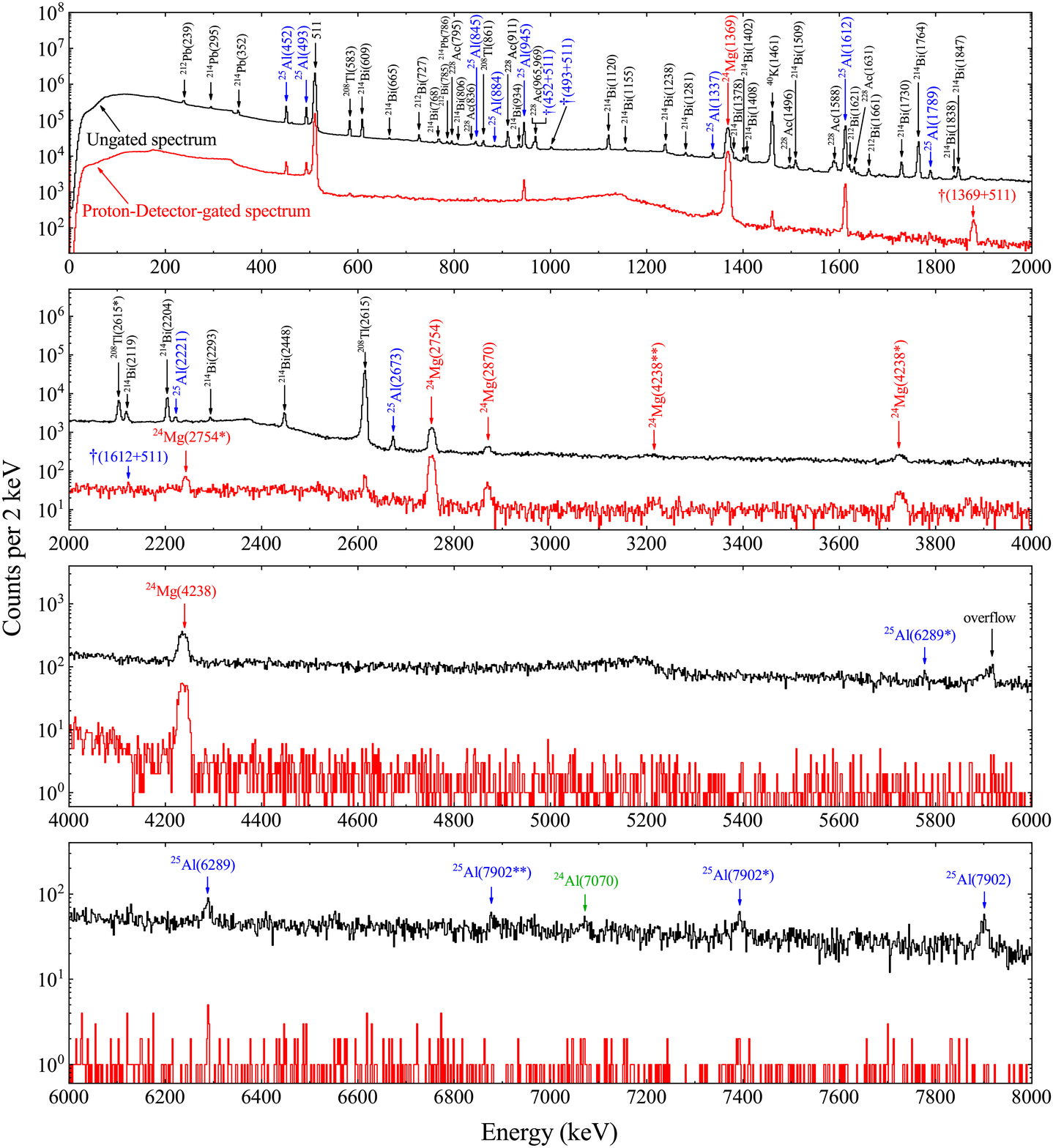}
\caption{\label{Fbg}$\gamma$-ray spectrum (upper; black) measured by the SeGA detectors showing the assignments for the photopeaks used to construct the $^{25}$Si decay scheme as well as those from room background. To reduce the room background contribution, a Proton-Detector-coincident $\gamma$-ray spectrum (lower; red) is produced by requiring coincidences with particle signals from the Proton Detector. Each photopeak is labeled by the emitting nucleus and its energy rounded to the closest integer in units of keV. Peaks labeled with one or two asterisks (*) correspond to single and double escape peaks, respectively. Peaks labeled with a single dagger ($\dagger$) are sum peaks with the summation noted. The bump at $\sim$5910~keV is caused by the overflow of one detector with unusual gain.}
\end{center}
\end{figure*}

Figure~\ref{Fbg} shows the full $\gamma$-ray spectrum measured by the SeGA detectors. The Proton-Detector-coincident $\gamma$-ray spectrum is also shown for comparison. This coincidence gate reduced the statistics for the $^{25}$Si($\beta p\gamma$)$^{24}$Mg peaks approximately by a factor of 4, which is related to the implant-decay cycle and the trigger efficiencies of the Proton Detector for protons. As can be seen from Fig.~\ref{Fbg}, the relative statistics for the $^{25}$Si($\beta\gamma$)$^{25}$Al peaks are even lower. This can be understood by considering the low trigger efficiency of the Proton Detector for $\beta$ particles. The coincidence condition suppresses the room background lines substantially and helps verify the origins of the $\gamma$-ray lines. Eight new $\beta$-delayed $\gamma$ rays are clearly observed in the $\beta$-decay of $^{25}$Si, and the results are summarized in Table~\ref{T25Al}. The uncertainty associated with the energy calibration of the SeGA detector and the statistical uncertainty from peak fitting were added in quadrature to obtain the total uncertainty of each $\gamma$ ray. For all the $^{25}$Si($\beta\gamma$)$^{25}$Al peaks reported in this paper, the dominant source of the $\gamma$-ray energy uncertainty is the statistical uncertainty. The absolute intensity of each $\gamma$ ray in the $\beta$ decay of $^{25}$Si is determined using the number of counts in the $\gamma$-ray peak, the $\gamma$-ray detection efficiency of the SeGA detectors, and the aforementioned total $\gamma$-ray intensity. A further correction for the summing effect is applied whenever necessary. The statistical uncertainty associated with the peak area is obtained from the peak-fitting procedure. The statistical uncertainty and the systematic uncertainties associated with the SeGA efficiency simulation are propagated through the calculation of each $\gamma$-ray intensity.

\begin{table*}\footnotesize
\caption{\label{T25Al}Results from the $^{25}$Si($\beta\gamma$)$^{25}$Al and $^{25}$Si($\beta p$)$^{24}$Mg decays obtained in the present paper. Columns 1$-$3 report the spin and parity ($J^\pi_i$), excitation energies ($E_i$), and $\beta$-feeding intensities ($I_\beta$) of each $^{25}$Al level populated by the $\beta$ decay of $^{25}$Si, respectively. Columns 4$-$7 report the excitation energies of the final states populated by $\gamma$-ray transitions ($E_f$) from each $^{25}$Al state, the laboratory frame energies of each $\gamma$-ray branch ($E_\gamma$), relative $\gamma$-ray branching ratios (B.R.), and $\gamma$-ray intensities per $^{25}$Si $\beta$ decay ($I_\gamma$), respectively. Columns 8$-$10 report the excitation energies ($E_f$) of the $^{24}$Mg states fed by proton emissions from each $^{25}$Al state, the energies of the emitted protons in the center-of-mass frame [$E_{p}(\mathrm{c.m.})$], and proton intensities ($I_p$) per $^{25}$Si $\beta$ decay, respectively.}
\begin{center}
\begin{ruledtabular}
\begin{tabular}{cccccccccc}
\multirow{2}[3]{*}{$J^\pi_i$~\cite{Firestone_NDS2009_25}} & \multirow{2}[3]{*}{$E_i$ (keV)} & \multirow{2}[3]{*}{$I_\beta$ (\%)\footnotemark[3]} & \multicolumn{4}{c}{$\gamma$-ray transition} & \multicolumn{3}{c}{Proton emission} \\
\cline{4-7}\cline{8-10}      &       &       & $E_f$ (keV) & $E_\gamma$ (keV) & B.R. (\%) & $I_\gamma$ (\%) & $E_f$ (keV) & $E_{p}(\mathrm{c.m.})$ (keV) & $I_p$ (\%)\footnotemark[16] \\
\hline
$5/2^+$ & 0     & 21.5(12)\footnotemark[4] &       &       &       &       &       &       &  \\
\cline{4-7}\cline{8-10}
$1/2^+$ & 451.7(5)\footnotemark[1] & $-$   & 0     & 451.7(5)\footnotemark[1] & 100   & 15.0(6) &       &       &  \\
\cline{4-7}\cline{8-10}
$3/2^+$ & 944.9(5)\footnotemark[1] & 22.6(7) & 452   & 493.3(7)\footnotemark[1] & 58.4(16) & 13.6(5) &       &       &  \\
      &       &       & 0     & 944.9(5)\footnotemark[1] & 41.6(16) & 9.7(4) &       &       &  \\
\cline{4-7}\cline{8-10}
$(7/2)^+$ & 1612.5(5)\footnotemark[1] & 15.2(9) & 0     & 1612.4(5)\footnotemark[1] & 100   & 15.3(9) &       &       &  \\
\cline{4-7}\cline{8-10}
$5/2^+$ & 1789.4(4) & 1.46(7) & 945   & 844.6(7) & 44(3) & 0.76(4) &       &       &  \\
      &       &       & 452   & 1337.4(16) & 30.6(20) & 0.52(3) &       &       &  \\
      &       &       & 0     & 1789.4(9) & 25.2(19) & 0.43(3) &       &       &  \\
\cline{4-7}\cline{8-10}
$3/2^+$ & 2673.4(4) & 6.8(15) & 1789  & 883.8(6) & 37(5) & 0.26(3) & 0     & 402.0(9)\footnotemark[6] & 6.1(15)\footnotemark[6] \\
      &       &       & 945 & 1728.3(8)\footnotemark[1] & 0.5(2)\footnotemark[1] & $-$\footnotemark[5]  &   &    &  \\
      &       &       & 452   & 2221.4(18) & 36(4) & 0.25(3) &       &       &  \\
      &       &       & 0     & 2673.6(6) & 26(3) & 0.184(17) &       &       &  \\
\cline{4-7}\cline{8-10}

$5/2^+$ & 3859.1(8)\footnotemark[1] & 0.30(16) & $-$ & $-$ & $-$ & $-$ & 0  & 1584(3)\footnotemark[7] & 0.30(16)\footnotemark[7] \\
\cline{4-7}\cline{8-10}

$3/2^+$ & 4192(4)\footnotemark[1] & 3.1(7) & $-$ & $-$ & $-$ & $-$ & 1369  & 554(10)\footnotemark[8] & 0.49(25)\footnotemark[8] \\
      &       &          &      &        &       &               & 0  & 1924.3(20)\footnotemark[7] & 2.6(7)\footnotemark[7] \\
\cline{4-7}\cline{8-10}

$5/2^+$ & 4582(2)\footnotemark[1] & 3.2(5) & $-$ & $-$ & $-$ & $-$ & 1369  & 943.7(11)\footnotemark[7] & 1.7(5)\footnotemark[7] \\
      &       &          &      &        &       &               & 0  & 2310.0(9)\footnotemark[7] & 1.5(3)\footnotemark[7] \\
\cline{4-7}\cline{8-10}

$(7/2)^+$ & 4906(4)\footnotemark[1] & 0.45(22) & $-$ & $-$ & $-$ & $-$ & 1369  & 1268(5)\footnotemark[9] & 0.41(22)\footnotemark[9] \\
      &       &          &      &        &       &               & 0  & 2632(10)\footnotemark[10] & 0.048(10)\footnotemark[10] \\
\cline{4-7}\cline{8-10}

$(3/2,5/2,7/2)^+$ & 5597(6)\footnotemark[1] & 0.5(3) & $-$ & $-$ & $-$ & $-$ & 0  & 3327(4)\footnotemark[7] & 0.5(3)\footnotemark[7] \\
\cline{4-7}\cline{8-10}

$(3/2,5/2,7/2)^+$ & 5804(4)\footnotemark[1] & 1.7(4) & $-$ & $-$ & $-$ & $-$ & 1369  & 2164(3)\footnotemark[7] & 1.7(4)\footnotemark[7] \\
\cline{4-7}\cline{8-10}

$-$ & 6063(7)\footnotemark[1] & 0.040(11) & $-$ & $-$ & $-$ & $-$ & 1369  & 2453(25)\footnotemark[11] & 0.040(11)\footnotemark[11] \\
\cline{4-7}\cline{8-10}

$(3/2,5/2,7/2)^+$ & 6170(2)\footnotemark[1] & 0.4(3) & $-$ & $-$ & $-$ & $-$ & 1369  & 2486(25)\footnotemark[11] & 0.10(3)\footnotemark[11] \\
      &       &          &      &        &       &               & 0  & 3896(8)\footnotemark[6] & 0.3(3)\footnotemark[6] \\
\cline{4-7}\cline{8-10}

$5/2^+$ & 6650(5)\footnotemark[1] & 0.42(25) & $-$ & $-$ & $-$ & $-$ & 1369  & 3006(11)\footnotemark[7] & 0.42(25)\footnotemark[7] \\
\cline{4-7}\cline{8-10}

$(3/2,5/2,7/2)^+$ & 6877(7)\footnotemark[1] & 0.42(16) & $-$ & $-$ & $-$ & $-$ & 1369  & 3236(6)\footnotemark[7] & 0.42(16)\footnotemark[7] \\
\cline{4-7}\cline{8-10}

$3/2^+$ & 6909(10)\footnotemark[1] & 0.035(11) & $-$ & $-$ & $-$ & $-$ & 0  & 4614(9)\footnotemark[10] & 0.035(11)\footnotemark[10] \\
\cline{4-7}\cline{8-10}

$3/2^+$ & 7118(5)\footnotemark[1] & 4.8(16) & $-$ & $-$ & $-$ & $-$ & 1369  & 3464(3)\footnotemark[7] & 3.6(15)\footnotemark[7] \\
      &       &          &      &        &       &               & 0  & 4845(4)\footnotemark[7] & 1.1(8)\footnotemark[7] \\
\cline{4-7}\cline{8-10}

$5/2^+$ & 7240(3)\footnotemark[1] & 1.3(5) & $-$ & $-$ & $-$ & $-$ & 4238  & 724(4)\footnotemark[12] & 0.036(15)\footnotemark[12] \\
      &       &          &      &        &       &               & 1369  & 3606(4)\footnotemark[7] & 1.0(5)\footnotemark[7] \\
      &       &          &      &        &       &               &  0      & 4980(4)\footnotemark[7] & 0.28(23)\footnotemark[7] \\
\cline{4-7}\cline{8-10}

$(7/2)^+$ & 7422(5)\footnotemark[1] & 0.16(6) & $-$ & $-$ & $-$ & $-$ & 4123  & 1037(16)\footnotemark[13] & 0.16(6)\footnotemark[13] \\
\cline{4-7}\cline{8-10}

$(3/2,5/2,7/2)^+$ & 7646\footnotemark[1] & 0.23(14) & $-$ & $-$ & $-$ & $-$ & 0  & 5382(11)\footnotemark[7] & 0.23(14)\footnotemark[7] \\
\cline{4-7}\cline{8-10}

$-$ & 7819(20)\footnotemark[1] & 0.32(9) & $-$ & $-$ & $-$ & $-$ & 0  & 5549(15)\footnotemark[11] & 0.32(9)\footnotemark[11] \\
\cline{4-7}\cline{8-10}

$5/2^+$ & 7902.0(14) & 13.4(16) & 1612  & 6289(3) & 36(10) & 0.086(20) & 4238  & 1380(5)\footnotemark[9] & 0.38(14)\footnotemark[9] \\
      &       &       & 945 & 6955(2)\footnotemark[1] & $<$15\footnotemark[5] & $<$0.037\footnotemark[5] & 4123  & 1492(6)\footnotemark[9] & 0.26(13)\footnotemark[9] \\
      &       &       & 0     & 7902(3) & 45(10) & 0.110(19) & 1369  & 4257(3)\footnotemark[7] & 10.3(14)\footnotemark[7] \\
      &       &       &       &       &       &       & 0     & 5628.8(15)\footnotemark[7] & 2.2(6)\footnotemark[7] \\
\cline{4-7}\cline{8-10}

$-$ & 7936(20)\footnotemark[1] & 0.45(13) & $-$ & $-$ & $-$ & $-$ & 1369  & 4345(17)\footnotemark[14] & 0.45(13)\footnotemark[14] \\
\cline{4-7}\cline{8-10}

$(3/2,5/2,7/2)^+$ & 8186(3)\footnotemark[1] & 1.0(4) & $-$ & $-$ & $-$ & $-$ & 4238  & 1684(12)\footnotemark[13] & 0.18(10)\footnotemark[13] \\
      &       &          &      &        &       &               & 4123  & 1794(3)\footnotemark[7] & 0.51(19)\footnotemark[7] \\
      &       &          &      &        &       &               &  1369  & 4551(5)\footnotemark[7] & 0.3(3)\footnotemark[7] \\
\cline{4-7}\cline{8-10}

$(3/2,5/2,7/2)^+$ & 9073(7)\footnotemark[1] & 0.13(10) & $-$ & $-$ & $-$ & $-$ & 0  & 6798(5)\footnotemark[15] & 0.13(10)\footnotemark[15] \\
\cline{4-7}\cline{8-10}

$-$ & 9275(25)\footnotemark[2] & 0.0127(17) & $-$ & $-$ & $-$ & $-$ & 0  & 7000(25)\footnotemark[2] & 0.0127(17)\footnotemark[2]  \\
\cline{4-7}\cline{8-10}

$-$ & 9415(30)\footnotemark[2] & 0.0127(17) & $-$ & $-$ & $-$ & $-$ & 0  & 7141(30)\footnotemark[2] & 0.0127(17)\footnotemark[2]  \\
\end{tabular}%
\end{ruledtabular}
\footnotetext[1]{\scriptsize Adopted from the data evaluation~\cite{Firestone_NDS2009_25}.}
\footnotetext[2]{\scriptsize Adopted from Ref.~\cite{Zhou_PRC1985}.}
\footnotetext[3]{\scriptsize Each $I_\beta$ is determined using the corresponding $I_\gamma$ in column 7 and $I_p$ in column 10. See Sec.~\ref{decay_scheme} for details.}
\footnotetext[4]{\scriptsize Average of the theoretical $I_\beta$ calculated using the USDC and USDI interactions.}
\footnotetext[5]{\scriptsize $\gamma$-ray branch indicated by the data evaluation~\cite{Firestone_NDS2009_25} but not observed in this paper due to limited sensitivity.}
\footnotetext[6]{\scriptsize Average of Refs.~\cite{Hatori_NPA1992,Robertson_PRC1993,Thomas_EPJA2004}.}
\footnotetext[7]{\scriptsize Average of Refs.~\cite{Sextro_Thesis1973,Hatori_NPA1992,Robertson_PRC1993,Thomas_EPJA2004}.}
\footnotetext[8]{\scriptsize Average of Refs.~\cite{Robertson_PRC1993,Thomas_EPJA2004}.}
\footnotetext[9]{\scriptsize Average of Refs.~\cite{Sextro_Thesis1973,Robertson_PRC1993,Thomas_EPJA2004}.}
\footnotetext[10]{\scriptsize Average of Refs.~\cite{Robertson_PRC1993,Hatori_NPA1992}.}
\footnotetext[11]{\scriptsize Adopted from Ref.~\cite{Robertson_PRC1993}.}
\footnotetext[12]{\scriptsize Observed in the present paper. See Sec.~\ref{25Sibp24Mg} for details.}
\footnotetext[13]{\scriptsize Average of Refs.~\cite{Sextro_Thesis1973,Robertson_PRC1993}.}
\footnotetext[14]{\scriptsize Average of Refs.~\cite{Sextro_Thesis1973,Hatori_NPA1992,Robertson_PRC1993}.}
\footnotetext[15]{\scriptsize Average of Refs.~\cite{Zhou_PRC1985,Hatori_NPA1992,Thomas_EPJA2004}.}
\footnotetext[16]{\scriptsize Except for the 724-keV proton branch, each $I_p$ is determined by taking a weighted average of available literature proton relative intensities and then normalized to the $^{25}$Si($\beta p\gamma$)$^{24}$Mg intensities determined in Sec.~\ref{25Sibpg24Mg}. The literature adopted for determining each $I_p$ is individually indicated in the corresponding footnote.}


\end{center}
\end{table*}

The 452- and 1612-keV $\gamma$ rays correspond to the 100\% transitions from the 452- and 1612-keV states to the $^{25}$Al ground state, respectively. The intensity of the 1612-keV $\gamma$ ray is corrected for the contribution of a nearby 1611.7-keV $\beta$-delayed $\gamma$ ray of $^{25}$Al~\cite{Firestone_NDS2009_25,Wilson_PRC1980}. There are two $\gamma$ rays which are emitted from the 945-keV state at 493 and 945~keV, and they are expected to have branching ratios of 61(4) and 39(4)\%, respectively~\cite{Piiparinen_ZP1972,Firestone_NDS2009_25}. We have improved these branching ratios to be 58.4(16) and 41.6(16)\% in this paper. The 1789-keV state is observed to be populated by the $\beta$ decay of $^{25}$Si for the first time. There are three $\gamma$ rays which are emitted from this state with energies of 844.6(7), 1337.4(16), and 1789.4(9)~keV, and their branching ratios are measured to be 44(3), 30.6(20), and 25.2(19)\%, respectively. The measured energies are consistent with the evaluated literature values of 844.6(7), 1337.8(7), and 1789.4(5)~keV~\cite{Firestone_NDS2009_25}. The three branching ratios are consistent with the literature values of 39.6(21), 36.1(18), and 23.3(10)\%, which are the weighted averages of five previous measurements with inflated uncertainty~\cite{Anyas-Weiss_CJP1969,van_Reenen_ZP1969,McCallum_CJP1971,Piiparinen_ZP1972,Tikkanen_PRC1991}. The excitation energy is determined to be 1789.2(6)~keV by combining the three $\gamma$-ray energies. This value is of comparable precision to the excitation energy of 1789.5(5)~keV reported in the data evaluation~\cite{Firestone_NDS2009_25}, and we have reevaluated the excitation energy to be 1789.4(4)~keV by taking a weighted average of the two values.

In all the previous $^{25}$Si decay measurements~\cite{Barton_CJP1963,McPherson_CJP1965,Hardy_CJP1965,Reeder_PR1966,Sextro_Thesis1973,Zhou_PRC1985,Garcia_PRC1990,Hatori_NPA1992,Robertson_PRC1993,Thomas_EPJA2004}, the proton-unbound states in $^{25}$Al were observed to decay only by proton emission. For the first time, we have observed the $\beta$-delayed $\gamma$ rays through two unbound $^{25}$Al states at 2673 and 7902~keV.

There are four known $\gamma$ rays which are emitted from the 2673-keV state at 883.8(8), 1728.3(8), 2221.5(8), 2673.1(6)~keV~\cite{Firestone_NDS2009_25}, and they are expected to have branching ratios of 42.8(8), 0.5(2), 31.4(7), and 25.3(5)\%, respectively~\cite{Powell_NPA1998}. We have observed three $\gamma$-ray branches from this state. As can be seen from Table~\ref{T25Al}, we have measured their energies and branching ratios to be 883.8(6)~keV~[37(5)\%], 2221.4(18)~keV~[36(4)\%], and 2673.6(6)~keV~[26(3)\%], respectively, which agree with the literature values~\cite{Firestone_NDS2009_25,Powell_NPA1998} at the 2$\sigma$ level. However, our sensitivity does not allow us to see the weakest 1728.3(8)-keV $\gamma$ ray from this state. The excitation energy is obtained to be 2673.4(5)~keV from the three $\gamma$-ray energies. Combining our result with the excitation energy of 2673.3(6)~keV from the data evaluation~\cite{Firestone_NDS2009_25} yields a weighted average of 2673.4(4)~keV.

The $5/2^+$ isobaric analog state (IAS) with isospin $T=3/2$ in $^{25}$Al is predicted to decay by 36 or 37 $\gamma$-ray branches by our shell-model calculations using the USDC or USDI Hamiltonian, respectively. The three most intense $\gamma$-ray branches at 6288(2), 6955(2), and 7900(2)~keV account for 92.4(15)\% of its total theoretical $\gamma$-ray branch. Their branching ratios measured by a $^{24}$Mg$(p,\gamma)^{25}$Al reaction experiment~\cite{Rogers_CJP1977,Firestone_NDS2009_25} are normalized to be 34(3), 12.0(19), and 46(3)\%, respectively. A theoretical percentage of 80.7(13)\% is used to normalize the branching ratios for the two $\gamma$ rays at 6289(3) and 7902(3)~keV observed in our work. Their branching ratios are determined to be 36(10) and 45(10)\%, respectively, in agreement with the previous measurement~\cite{Rogers_CJP1977}. The highest-energy $\gamma$ ray at 7902(3)~keV is assigned as the deexcitation from the IAS to the ground state of $^{25}$Al. Its single escape and double escape peaks are also observed, and the excitation energy of the IAS is determined to be 7903(2)~keV by combining the full photopeak energy and escape-peak energies. This value is of comparable precision to the excitation energy of 7901(2)~keV reported in the data evaluation~\cite{Firestone_NDS2009_25}, and we have reevaluated the excitation energy of the IAS to be 7902.0(14)~keV by taking a weighted average of the two values. The weakest 6955-keV $\gamma$-ray line is less than 3$\sigma$ above the background level in our spectrum, and we estimate the 90\% confidence upper limit for its branching ratio to be $<$15\%, in agreement with the literature value of 12.0(19)\%~\cite{Rogers_CJP1977,Firestone_NDS2009_25}.

\subsection{\label{25Sibp24Mg}$^{25}$Si($\beta p$)$^{24}$Mg}
\begin{figure}
\begin{center}
\includegraphics[width=8.6cm]{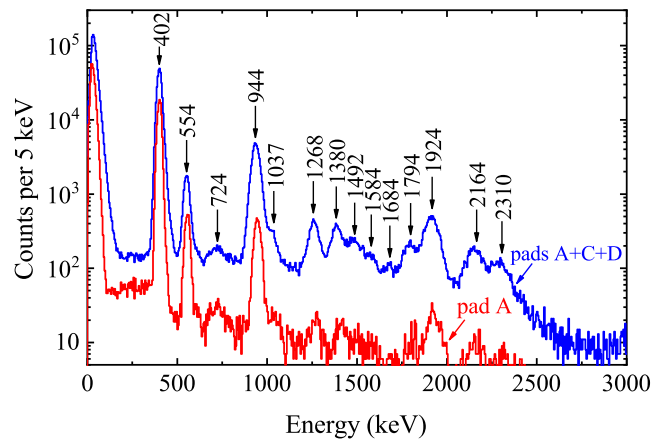}
\caption{\label{Fbp}Proton spectra measured by three central pads A+C+D (upper; blue) and central pad A (lower; red). Each proton peak from the $\beta$-delayed proton decay of $^{25}$Si is labeled with its center-of-mass energy rounded to the closest integer in units of keV.}
\end{center}
\end{figure}

The $\beta$-delayed proton spectrum of $^{25}$Si is shown in Fig.~\ref{Fbp}. The event-level summing of three central pads (A+C+D) and an individual spectrum for pad A are shown for comparison. The single pad spectrum is generated with anticoincidence cuts on all other pads, resulting in a lower background and a fast-declining efficiency as a function of proton energy. Robertson~\textit{et al}.~\cite{Robertson_PRC1993} observed 13 proton peaks below 2310~keV. Hatori~\textit{et al}.~\cite{Hatori_NPA1992} did not observe six of them, and Thomas~\textit{et al}.~\cite{Thomas_EPJA2004} did not observe two of them at 1037 and 1684~keV. In the present work, all 13 known proton peaks below 2310~keV have been observed. We have reevaluated the proton energies by taking a weighted average of available literature proton energies with inflated uncertainty. The proton intensities are reevaluated by taking a weighted average of available literature proton relative intensities and then normalized to the $^{25}$Si($\beta p\gamma$)$^{24}$Mg intensities determined in Sec.~\ref{25Sibpg24Mg}. A total of 34 proton energies and intensities are evaluated and listed in Table~\ref{T25Al}. The uncertainties of proton intensities reported by Robertson~\textit{et al}.~\cite{Robertson_PRC1993} were unrealistically small~\cite{Batchelder_ADNDT2020}, and therefore, we take an unweighted average of literature relative intensities and assign an uncertainty that covers all literature central values. For the three proton emissions at 2453, 2486, and 5549~keV only observed by Robertson~\textit{et al}.~\cite{Robertson_PRC1993}, the uncertainties evaluated in this way become zero. Hence, we extract the residuals between the averaged literature relative intensities and those measured by Robertson~\textit{et al}.~\cite{Robertson_PRC1993} based on all other proton emissions. We derive a standard deviation of all the residuals, and this standard deviation is then factored into the uncertainties of the intensities for the 2453-, 2486-, and 5549-keV protons. As shown in Fig.~\ref{Fbp}, the $\beta$-particle background in our proton spectrum is suppressed to as low as 100~keV, enabling the clear identification of a new proton peak at 724(4)~keV. We derive a detection efficiency curve for all other protons based on the number of counts in each peak observed in the proton spectrum and its corresponding intensity. We then interpolate the efficiency at 724~keV and determine the intensity for the 724-keV proton emission to be 0.036(15)\%.


\begin{figure}
\begin{center}
\includegraphics[width=8.6cm]{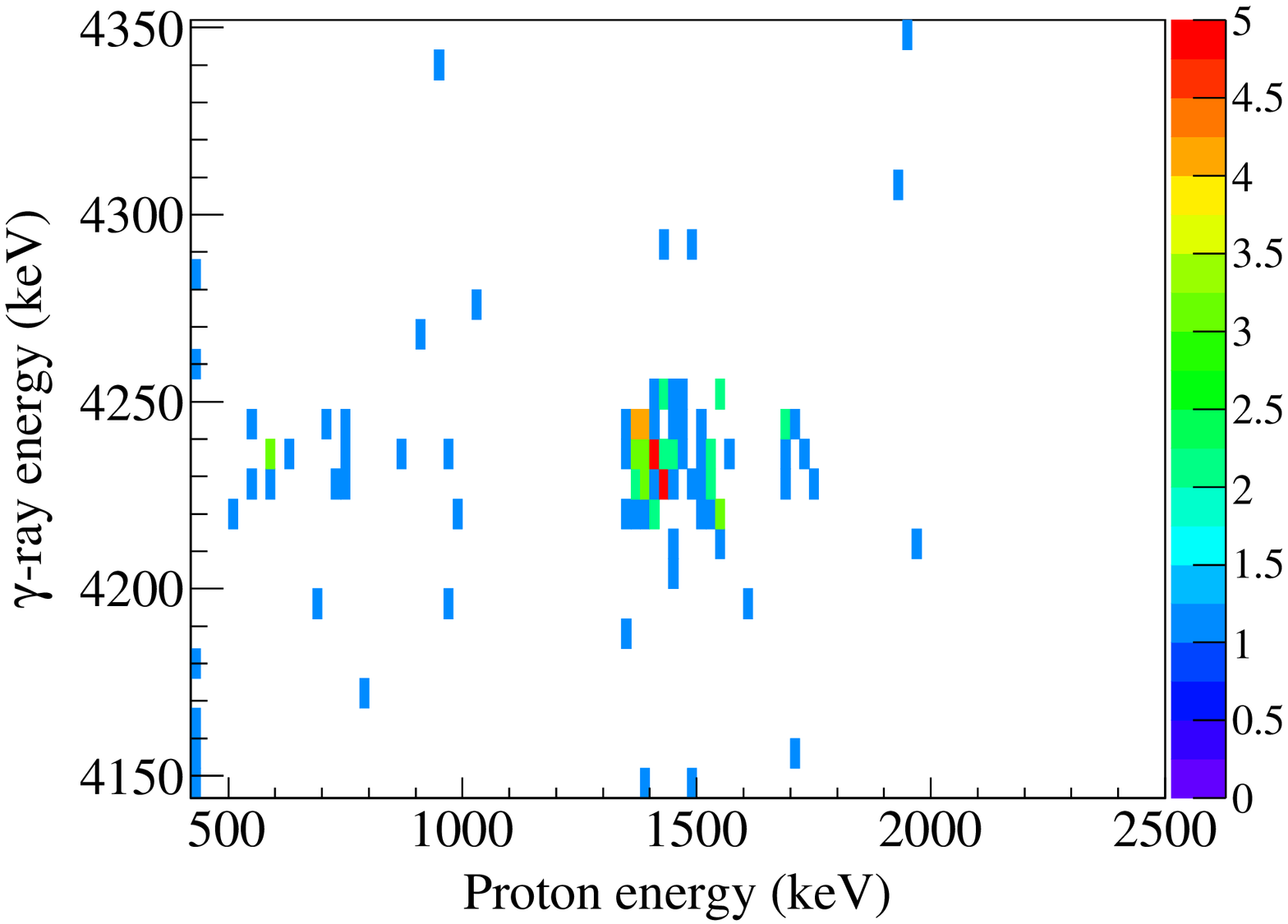}
\includegraphics[width=8.6cm]{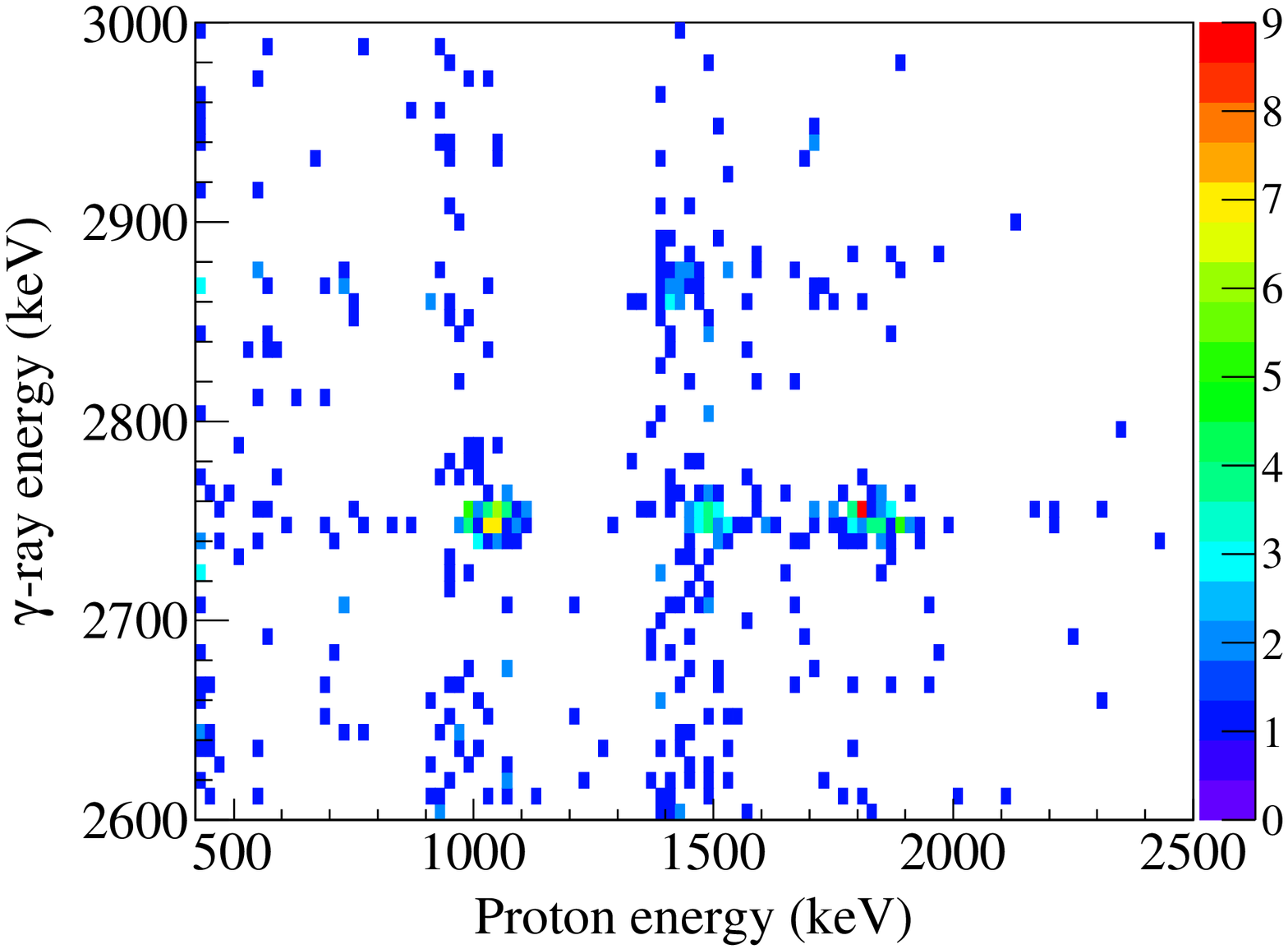}
\includegraphics[width=8.6cm]{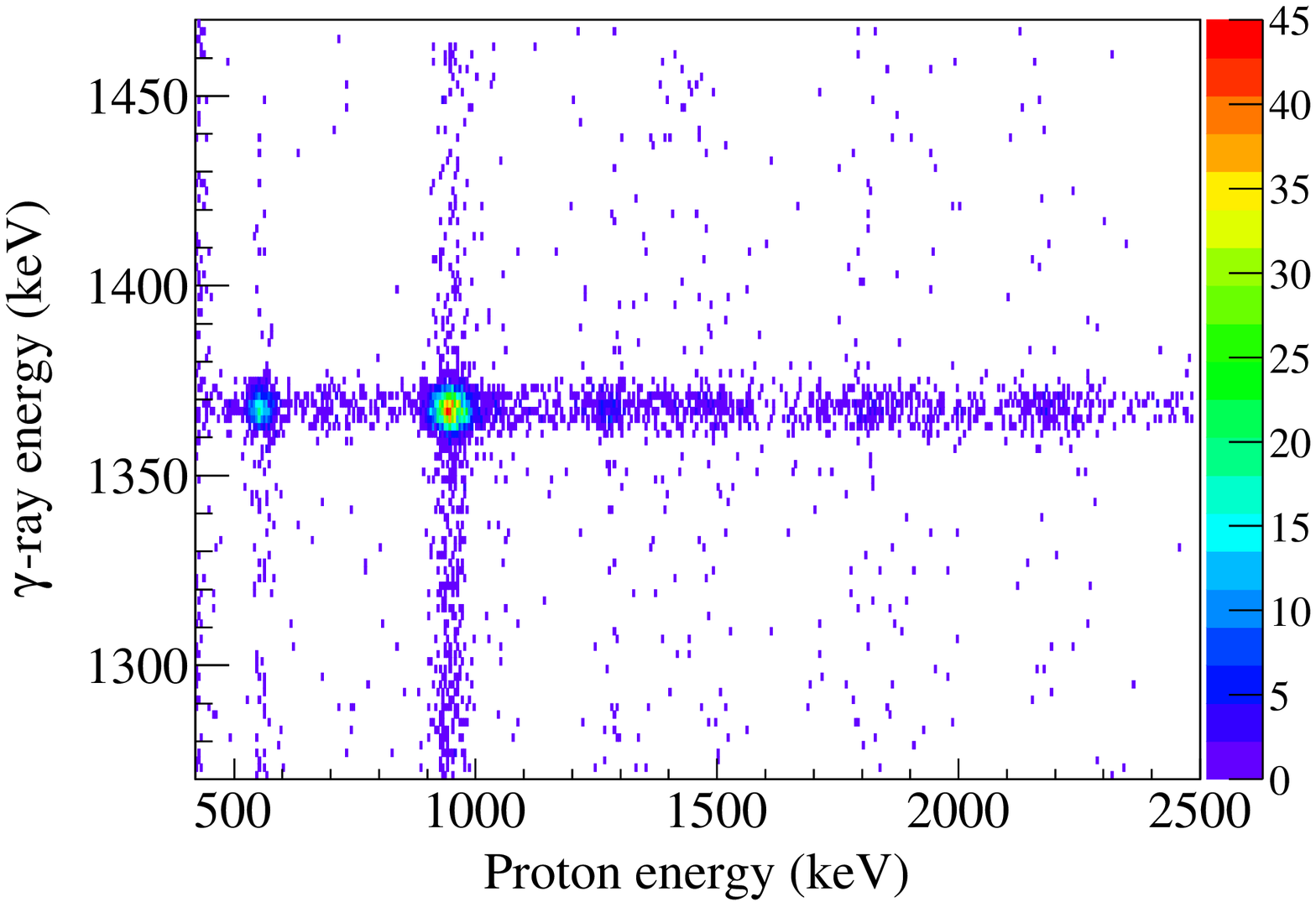}
\caption{\label{Fgp}Coincidence spectrum between the Proton Detector and SeGA detection for $^{25}$Si decay. The $\gamma$-ray spectrum is magnified at 4238~keV (top panel), 2754 and 2870~keV (middle panel), and 1369~keV (bottom panel), corresponding to the four $\gamma$ rays originating from the three lowest-lying $^{24}$Mg states.}
\end{center}
\end{figure}

\begin{table}
\caption{\label{Tgp}Coincidence matrix of the protons and $\gamma$ rays measured in the $\beta$ decay of $^{25}$Si. The first row corresponds to the $\gamma$-ray energy on which the gate is set. The following rows indicate the protons observed in the gated spectrum. Protons observed in coincidence are indicated with a checkmark (\checkmark) if the signal is statistically significant. All the energies are rounded to the closest integer and are given in units of keV. }
\begin{center}
\begin{ruledtabular}
\begin{tabular}{ccccc}
        & 1369  & 2754  & 2870  & 4238 \\
\hline
402    &                   &                   &                   &                    \\
554    & \checkmark &                   &                   &                    \\
724    &                   &                   & \checkmark & \checkmark  \\
944    & \checkmark &                   &                   &                    \\
1037  & \checkmark & \checkmark &                   &                    \\
1268  & \checkmark &                   &                   &                    \\
1380  & \checkmark &                   & \checkmark & \checkmark  \\
1492  & \checkmark & \checkmark &                   &                    \\
1584  &                   &                   &                   &                    \\
1684  &                   &                   & \checkmark & \checkmark  \\
1794  & \checkmark & \checkmark &                   &                    \\
1924  &                   &                   &                   &                    \\
2164  & \checkmark &                   &                   &                    \\
2310  &                   &                   &                   &                    \\
\end{tabular}%
\end{ruledtabular}
\end{center}
\end{table}

\subsection{\label{decay_scheme}Proton-$\gamma$ coincidences and decay scheme}

\begin{figure*}
\begin{center}
\includegraphics[width=17cm]{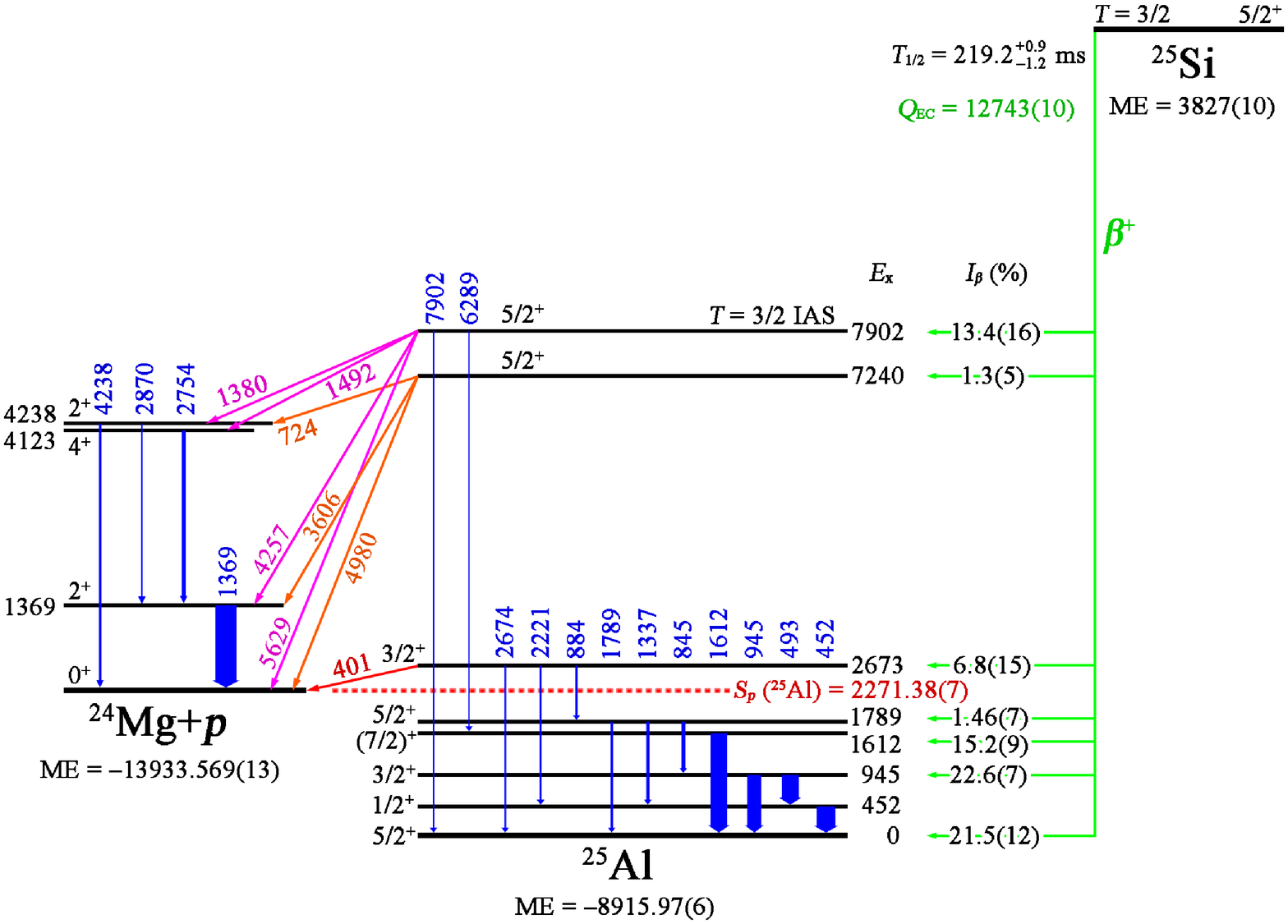}
\caption{\label{Scheme}Simplified decay scheme of $^{25}$Si. The mass excesses, separation energies, $Q$ values, spins, and parities are adopted from the data evaluations~\cite{Wang_CPC2017,Firestone_NDS2009_25,Firestone_NDS2007_24}. The half-life is the weighted average of Refs.~\cite{McPherson_CJP1965,Reeder_PR1966,Hatori_NPA1992,Robertson_PRC1993} and the present paper. The $\gamma$-ray energies and the excitation energies deduced from these $\gamma$-ray energies are rounded to the nearest keV. Each $\gamma$-ray transition is denoted by a vertical arrow followed by its $\gamma$-ray energy, and the corresponding $\gamma$-ray transition intensity is denoted by the thicknesses of the arrow. Each $\beta$-decay transition is depicted by an arrow on the right side of the figure followed by its feeding intensity. The 2673- and 7902-keV $^{25}$Al states are observed to decay by both proton and $\gamma$-ray emissions. The newly observed 724-keV proton is emitted from the 7240-keV $^{25}$Al state. Each proton transition is denoted by an arrow between its initial and final states labeled alongside by its center-of-mass energy. For the sake of brevity, we omit other unbound $^{25}$Al states. All the energies and masses are given in units of keV. See Table~\ref{T25Al} for details.}
\end{center}
\end{figure*}

In order to reliably construct the decay scheme, it is desirable to conduct a $p\gamma$ coincidence analysis. Only two previous measurements showed a handful of $p\gamma$ coincidences. Garc\'{i}a~\textit{et al}. reported the coincidences between the 1369-keV $\gamma$ ray and 3464-, 3606-, and 4257-keV protons~\cite{Garcia_PRC1990}. Thomas~\textit{et al}. confirmed the 4257-1369 $p\gamma$ coincidence~\cite{Thomas_EPJA2004}. In the present work, many more $p\gamma$ coincidences have been directly observed. Figure~\ref{Fgp} shows three regions of the two-dimensional coincidence spectrum between the protons and $\gamma$ rays from $^{25}$Si decay. The protons and $\gamma$ rays detected in coincidence are summarized in Table~\ref{Tgp} in the form of a coincidence matrix.

The newly identified 724-keV proton is observed in coincidence with the 2870- and 4238-keV $\gamma$ rays. Hence, we assign it as a proton transition to the 4238-keV excited state of $^{24}$Mg and obtain an excitation energy of $E_x=7234(4)$~keV for its proton-emitting state in $^{25}$Al. The excitation energy is consistent with a $5/2^+$ proton-emitting state, which was previously measured to be 7240(7)~keV~\cite{Browne_PRC1973}, 7240(3)~keV~\cite{Prior_NPA1991}, and 7239(5)~keV~\cite{Fujita_PRL2004}, 7243(12)~keV~\cite{Sextro_Thesis1973}, 7248(5)~keV~\cite{Hatori_NPA1992}, 7245(8)~keV~\cite{Robertson_PRC1993}, and 7255(7)~keV~\cite{Thomas_EPJA2004}, respectively. A decay width of 19(4)~keV was reported in a polarized proton scattering experiment~\cite{Prior_NPA1991}, which explains the broad peak shape at 724~keV observed in our proton spectrum. Previous $^{25}$Si decay experiments~\cite{Sextro_Thesis1973,Hatori_NPA1992,Robertson_PRC1993,Thomas_EPJA2004} observed two proton peaks at 3606 and 4980~keV, corresponding to the proton transitions from this state to the first excited state and ground state of $^{24}$Mg, respectively. Our Proton Detector is not sensitive to those high-energy protons. Hence, we have determined the $I_p=1.0(5)\%$ and 0.28(23)\% for the 3606- and 4980-keV proton branches, respectively, based on the relative proton intensities measured in previous $^{25}$Si decay experiments~\cite{Sextro_Thesis1973,Hatori_NPA1992,Robertson_PRC1993,Thomas_EPJA2004} and the aforementioned proton feedings to each $^{24}$Mg state (Table~\ref{T24Mg}). No $\gamma$-ray branches populating or deexciting the 7240-keV state have been observed; therefore, the $\beta$ feeding of the 7240-keV state is determined by adding up the intensities of the three proton branches from this state.

The $\gamma$-ray transitions are placed in the decay scheme shown in Fig.~\ref{Scheme} based on the known level scheme in the database~\cite{Firestone_NDS2009_25,Firestone_NDS2007_24}, as well as including consideration of spin and parity selection rules and the $\gamma$-ray energy relationships. The level scheme is also verified using $\gamma\gamma$ coincidences. Except for the $\gamma\gamma$ coincidences associated with the two relatively weak $\gamma$ rays originating from the IAS, all the expected $\gamma\gamma$ coincidences between other $\gamma$-ray transitions are observed in our paper. All the bound states of $^{25}$Al are observed to be populated in the $\beta$ decay of $^{25}$Si. The $\beta$-feeding intensity to a $^{25}$Al bound state is determined by subtracting the intensity of the $\gamma$ rays feeding this level from the intensity of the $\gamma$ rays deexciting this level. The feeding of the first excited state of $^{25}$Al with $J^\pi=1/2^+$ is consistent with its population by $\beta\gamma$ decay rather than directly by a second-forbidden $\beta$ transition. It is possible that there exist weak, unobserved $\gamma$ feedings from high-lying states, and the apparent $\beta$ feedings for low-lying states are thus higher than the true $\beta$ feedings due to the pandemonium effect~\cite{Hardy_PLB1977}. We have assessed the extent of this effect based on the shell-model calculations, and the unobserved $\gamma$ feedings from high-lying states to each low-lying state are expected to be negligible ($<$$10^{-4}$) due to the dominance of proton emission. The $\beta$-feeding intensity to a $^{25}$Al unbound state is determined by adding up the intensities of the proton and $\gamma$-ray branches from this state. The excitation energies and $\beta$-feeding intensities of all the $^{25}$Al levels populated by $^{25}$Si $\beta$ decay are tabulated in Table~\ref{T25Al}.

\subsection{\label{Shell-model calculations}Shell-model calculations}
We have performed the theoretical calculations using the shell-model code \textsc{NuShellX}~\cite{Brown_NDS2014} in the $sd$-shell-model space involving the $\pi0d_{5/2}$, $\pi1s_{1/2}$, $\pi0d_{3/2}$, $\nu0d_{5/2}$, $\nu1s_{1/2}$, and $\nu0d_{3/2}$ valence orbits. Two modified universal \textit{sd} (USD) Hamiltonian~\cite{Brown_ARNPS1988}, USDC and USDI, which directly incorporate Coulomb and other isospin-breaking interactions~\cite{Magilligan_PRC2020} were used. A quenching factor $q=0.6$ for the Gamow-Teller strength was used in our theoretical calculation based on the average over the whole $sd$ shell. Given the quenching factors in $sd$ shell ranging from 0.5 near $^{40}$Ca to 0.7 near $^{16}$O, the theoretical uncertainty associated with the $A=25$, $q=0.6$ is estimated to be $\pm0.1$. The theoretical log~$ft$ and $B$(GT) values are reported in Table~\ref{Tmirror}. In general, the characteristics of the decay scheme measured in the present paper including the excitation energies, $\beta$-feeding intensities, log~$ft$, $B$(GT), and $\gamma$-ray and proton partial widths for the states of $^{25}$Al can be reproduced well within the framework of the nuclear shell model.

Low-lying states and the $T=3/2$ IAS in $^{25}$Al have been unambiguously identified and their excitation energies have been well measured. Given that decay widths and intensities are very sensitive to energies, we have applied a correction to the theoretical $\beta$ feedings, $\gamma$-ray partial widths ($\Gamma_\gamma$), and proton partial widths ($\Gamma_p$) based on the experimental energies. The $I_\beta=T/t$ is determined using the half-life of $^{25}$Si, $T$, and the individual partial half-life for each transition, $t$. The latter is scaled from the theoretical $t$ by the phase space factor $f$ using the experimental $\beta$-decay energy of $^{25}$Si and the excitation energy of each $^{25}$Al state under the assumption of constant $ft$ value. Each theoretical $\Gamma_p$ is calculated by using the theoretical spectroscopic factor and the barrier-penetration factor~\cite{Barnett_CPC1982} corrected for the experimental resonance energy. Each theoretical $\Gamma_\gamma$ is obtained using the effective $M$1 and $E$2 transition operators from Ref.~\cite{Richter_PRC2008} and then scaled for the $E^{2L+1}_\gamma$ energy dependence, where $L$ denotes the multipolarity of the radiation.


\begin{table*}
\caption{\label{Tmirror}Comparison of the mirror transitions in $^{25}$Si and $^{25}$Na $\beta$ decays. Column 1 lists the excitation energies of each $^{25}$Al state. Columns 2 and 3 report the log~$ft$ and $B$(GT) values for the each $^{25}$Si $\beta$ transition. Column 4 shows the $J^\pi$ assignments~\cite{Firestone_NDS2009_25}. Columns 5$-$7 list the results of the mirror $^{25}$Na $\beta$-decay transitions. The mirror-asymmetry parameters $\delta$ are reported in the last column. The USDC and USDI shell-model calculated results for both $^{25}$Si and $^{25}$Na decays are shown for comparison. }
\begin{center}
\begin{ruledtabular}
\begin{tabular}{cccccccc}
\multicolumn{3}{c}{$^{25}$Si$\rightarrow^{25}$Al Present experiment}  &  & \multicolumn{3}{c}{$^{25}$Na$\rightarrow^{25}$Mg~\cite{Alburger_PRC1971,Alburger_NPA1982}}  & \\
 \cline{1-3}   \cline{5-7}
$^{25}$Al $E_x$ (keV)  & log~$ft$ & $B$(GT)  & $J^\pi$~\cite{Firestone_NDS2009_25}    & $^{25}$Mg $E_x$ (keV)  & log~$ft$ & $B$(GT) &  $\delta$ \\
\hline
0             & 5.306(25) & 0.0187(11) &  $5/2^+$ &    0               & 5.251(15) & 0.0212(7)    & 0.14(8) \\
944.9(5)  & 5.109(14)  &  0.0294(9)  & $3/2^+$ & 974.749(24)  &  5.043(6)  & 0.0342(5)    & 0.16(4) \\
1612.5(5) &  5.15(3)   &  0.0269(19)  & $7/2^+$  & 1611.772(11) & 5.030(8) & 0.0352(7)   & 0.31(9) \\
1789.4(4) & 6.131(22) & 0.00279(14) & $5/2^+$ & 1964.620(24) & 6.045(9) & 0.00340(7) & 0.22(7) \\
2673.4(4) & 5.27(10)   &  0.020(5)    &  $3/2^+$ & 2801.46(3)  & 5.246(9)   &  0.0214(5)   & 0.05(24) \\
\hline
\multicolumn{3}{c}{$^{25}$Si$\rightarrow^{25}$Al USDC}  &  & \multicolumn{3}{c}{$^{25}$Na$\rightarrow^{25}$Mg USDC}  & \\
 \cline{1-3}   \cline{5-7}
$^{25}$Al $E_x$ (keV)  & log~$ft$ & $B$(GT)  &      & $^{25}$Mg $E_x$ (keV)  & log~$ft$ & $B$(GT) &  $\delta$ \\
\hline
      0 & 5.314 & 0.0183   & $5/2^+$ &      0 & 5.312 & 0.0184 & 0.01 \\
1015 & 5.093 & 0.0305   & $3/2^+$ & 1072 & 5.087 & 0.0309 & 0.01 \\
1723 & 5.208 & 0.0234   & $7/2^+$ & 1708 & 5.183 & 0.0248 & 0.06 \\
1882 & 6.070 & 0.0032   & $5/2^+$ & 2012 & 6.140 & 0.0027 & $-$0.15 \\
2739 & 5.127  & 0.0282  & $3/2^+$ & 2834 & 5.135 & 0.0277 & $-$0.02 \\
\hline
\multicolumn{3}{c}{$^{25}$Si$\rightarrow^{25}$Al USDI}  &  & \multicolumn{3}{c}{$^{25}$Na$\rightarrow^{25}$Mg USDI}  & \\
 \cline{1-3}   \cline{5-7}
$^{25}$Al $E_x$ (keV)  & log~$ft$ & $B$(GT)  &      & $^{25}$Mg $E_x$ (keV)  & log~$ft$ & $B$(GT) &  $\delta$ \\
\hline
     0 & 5.293 & 0.0192    & $5/2^+$ &      0 & 5.291 & 0.0193 & 0.01 \\
1013 & 5.113 & 0.0291   & $3/2^+$ & 1068 & 5.107 & 0.0295 & 0.01 \\
1722 & 5.206 & 0.0235   & $7/2^+$ & 1707 & 5.181 & 0.0249 & 0.06 \\
1890 & 6.084 & 0.0031   & $5/2^+$ & 2020 & 6.173 & 0.0025 & $-$0.18 \\
2761 & 5.139 & 0.0274   & $3/2^+$ & 2854 & 5.147 & 0.0269 & $-$0.02 \\
\end{tabular}
\end{ruledtabular}
\end{center}
\end{table*}

\subsection{Mirror asymmetry}
With the $\beta$-decay energy of $^{25}$Si $Q_{\mathrm{EC}}(^{25}$Si$)=12743(10)$~keV~\cite{Wang_CPC2017}, the $^{25}$Si half-life of $219.2^{+0.9}_{-1.2}$~ms, the excitation energies of $^{25}$Al states, and the $\beta$-feeding intensities to $^{25}$Al states measured in the present paper, the corresponding log~$ft$ values for each $^{25}$Al state can be calculated through the \textsc{logft} analysis program provided by the National Nuclear Data Center website~\cite{Gove_ADNDT1971}. The corresponding Gamow-Teller transition strengths, $B$(GT), are calculated from the $ft$ values using the following relation:
\begin{equation}
ft=\frac{K}{g_V^2B(\text{F})+g_A^2B(\text{GT})},
\end{equation}
where $K/g_V^2=6144.48\pm3.70$~s~\cite{Hardy_PRC2020} and $(g_A/g_V)^2=(-1.2756\pm0.0013)^2$~\cite{Zyla_PTEP2020}, with $g_V$ and $g_A$ being the free vector and axial-vector coupling constants of the weak interaction. Our shell-model calculations predict that the Fermi transition strengths $B$(F) are negligible for low-lying $^{25}$Al states.

The degree of isospin-symmetry breaking can be quantified by the mirror-asymmetry parameter $\delta=ft^+/ft^--1$, where the $ft^+$ and $ft^-$ values are associated with the $\beta^{+}$ decay of $^{25}$Si and the $\beta^{-}$ decay of $^{25}$Na, respectively. $\delta=0$ denotes perfect isospin symmetry. The log~$ft$ and $B$(GT) values for each $\beta$-decay transition and the corresponding mirror-asymmetry parameter are summarized in Table~\ref{Tmirror}. Limited by the $Q_{\mathrm{\beta-}}=3835.0(12)$~keV, only five $^{25}$Mg states were observed to be populated by $^{25}$Na $\beta$ decay~\cite{Jones_PRC1970,Alburger_PRC1971,Alburger_NPA1982}, and each one of them can be matched with a specific $^{25}$Al state measured in our paper. Thomas~\textit{et al}.~\cite{Thomas_EPJA2004} compared four transitions between the mirror nuclei $^{25}$Si and $^{25}$Na, and their mirror-asymmetry parameters for three bound states are consistent with but less precise than our values. We did not observe mirror asymmetry between the transitions to the second $3/2^+$ state. We observed some small but significant asymmetries for the other four low-lying states. The theoretical $B$(GT) values for $^{25}$Si decay are in agreement with our experimental values considering the theoretical uncertainties. Our shell-model calculations somewhat underestimated the $B$(GT) value for the $7/2^+$ state but slightly overestimated that for the second $3/2^+$ state compared with $^{25}$Na $\beta$-decay measurements~\cite{Jones_PRC1970,Alburger_PRC1971,Alburger_NPA1982}, suggesting that a more careful theoretical treatment is needed, e.g., using the shell model in conjunction with more realistic radial wave functions and sums over parentages in the $A-1$ nuclei, including a change in the radial wave function overlap factors and how this is connected to the Thomas-Ehrman shifts. These calculations call for more theoretical efforts in the future.

\subsection{$^{25}$Al 2673-keV state}
The $\beta$ feeding of the 2673-keV state of $^{25}$Al is measured to be $I_{\beta}=6.8(15)\%$ by the sum of the intensities of the four $\beta$-delayed $\gamma$ rays deexciting the 2673-keV state $I_{\gamma}=0.70(4)\%$ and the intensity of the 402-keV proton $I_{p}=6.1(15)\%$. The $\beta$ feeding of the 2673-keV state is in agreement with the previous measured values of $I_{\beta}=6.93(76)\%$~\cite{Hatori_NPA1992} and  $I_{\beta}=8.2(15)\%$~\cite{Robertson_PRC1993}. Thomas~\textit{et al}. reported a smaller $I_{\beta}=4.8(3)\%$~\cite{Thomas_EPJA2004}, in which the 2673-keV state was assumed to decay only via proton emission. The ratio $I_{\gamma}/I_{p}$ is equal to the ratio $\Gamma_\gamma/\Gamma_p$. We determine an experimental value of $\Gamma_\gamma/\Gamma_p=0.11(3)$, in agreement with the $\Gamma_\gamma/\Gamma_p=0.143(16)$ derived from a $^{24}$Mg$(p,\gamma)^{25}$Al reaction measurement~\cite{Powell_NPA1999}.

\begin{table}\footnotesize
\caption{\label{T2673}Decay properties of the 2673-keV $3/2^+$ state in $^{25}$Al.}
\begin{center}
\begin{ruledtabular}
\begin{tabular}{cccccc}
Reference & $\Gamma_\gamma$~(meV) & $\Gamma_p$~(meV) & $\Gamma_\gamma/\Gamma_p$ & $\omega\gamma$~(meV) & $\tau$~(fs) \\
\hline
Refs.~\cite{Powell_NPA1998,Powell_NPA1999} & 23.8(15) & 166(16) & 0.143(16) & 41.6(26) & $6.1^{+4.8}_{-3.7}$ \\
USDC & 20.6 & 173 & 0.119 & 36.8 & 3.4 \\
USDI  & 21.2 & 173 & 0.123 & 37.7 & 3.4 \\
Present & 23(8)\footnotemark[1]  & 202(48)\footnotemark[1]  & 0.11(3)  & 41.6(26)\footnotemark[2]  & 2.9(7)  \\
\end{tabular}%
\end{ruledtabular}
\footnotetext[1]{Deduced from the $\omega\gamma$ measured by Refs.~\cite{Powell_NPA1998,Powell_NPA1999} and the $I_\gamma$ and $I_p$ measured in the present paper.}
\footnotetext[2]{Adopted from Refs.~\cite{Powell_NPA1998,Powell_NPA1999}.}
\end{center}
\end{table}

Another $^{24}$Mg$(p,\gamma)^{25}$Al reaction measurement determined the resonance strength of the 2673-keV state to be $\omega\gamma=41.6(26)$~meV~\cite{Powell_NPA1998}. The total decay width $\Gamma_{\mathrm{tot}}$ is the sum of the $\Gamma_p$ and $\Gamma_\gamma$ since they represent the only two open decay channels for the 402-keV resonance in $^{25}$Al. The $\Gamma_{\mathrm{tot}}$ and $\omega\gamma$ are related by the following expression:

\begin{equation}
\omega\gamma=\frac{2J_r+1}{(2J_p+1)(2J_{T}+1)}\frac{\Gamma_p\times\Gamma_\gamma}{\Gamma_\mathrm{tot}},
\end{equation}

where $J_r=3/2$ is the spin of the 402-keV resonance, $J_p=1/2$ is the spin of the proton, and $J_T=0$ is the spin of the ground state of $^{24}$Mg. The lifetime of the 2673-keV $^{25}$Al state was previously measured to be $\tau=6.1^{+4.8}_{-3.7}$~fs using the Doppler shift attenuation method~\cite{Powell_NPA1999}. This value was converted to a half-life of 4(3)~fs and adopted by the evaluation~\cite{Firestone_NDS2009_25}. The lifetime is inversely proportional to the decay width by $\tau=\hbar/\Gamma_{\rm{tot}}$, where $\hbar$ is the Planck constant. Combining the branching ratio $\Gamma_\gamma/\Gamma_p$ measured in this paper with the literature $\omega\gamma$ value yields a lifetime for the 2673-keV state of 2.9(7)~fs, which is consistent with, as well as more precise than, the previously measured lifetime~\cite{Powell_NPA1999}. The decay properties of the 2673-keV state in $^{25}$Al are summarized and compared to the USDC and USDI shell-model calculations in Table~\ref{T2673}, and good agreement is obtained for all the quantities.

\subsection{$^{25}$Al $T=3/2$ IAS at 7902~keV}
The proton partial width of the lowest $T=3/2$ IAS in $^{25}$Al was determined to be $\Gamma_p=155(50)$~eV~\cite{Ikossi_NPA1976,Ikossi_PRL1976} and 105(18)~eV~\cite{Wilkerson_NPA1992}, respectively, in two proton scattering measurements with polarized-proton beams. These two results agree, and a weighted average $\Gamma_p$ is deduced to be 111(17)~eV. The $\gamma$-ray partial width of the IAS was previously determined to be $\Gamma_\gamma=2.0(10)$~eV in a $^{24}$Mg$(p,\gamma)^{25}$Al reaction yield measurement~\cite{Morrison_PR1968} by adopting a proton branching ratio $I_{p0}/I_{p\mathrm{tot}}=0.17$ from the $^{25}$Si $\beta$-delayed proton measurement~\cite{Reeder_PR1966}. $I_{p0}$ is the intensity of the proton emission from the IAS proceeding to the ground state of $^{24}$Mg. $I_{p\mathrm{tot}}$ is the total intensity of the proton branches of the IAS. However, another $^{24}$Mg$(p,\gamma)^{25}$Al reaction study~\cite{Rogers_CJP1977} reported a much smaller $\Gamma_\gamma=0.50(13)$~eV based on the measured resonance strength $\omega\gamma=0.25(6)$~eV and the proton branching ratio $I_{p0}/I_{p\mathrm{tot}}=0.168(13)$ from another $^{25}$Si $\beta$-delayed proton measurement~\cite{Sextro_Thesis1973}. The ratio of the $\gamma$-ray partial width to the proton partial width is deduced to be either $\Gamma_\gamma/\Gamma_p=0.018(9)$ by adopting the $\Gamma_\gamma$ of Ref.~\cite{Morrison_PR1968} or $\Gamma_\gamma/\Gamma_p=0.0045(13)$ by adopting the $\Gamma_\gamma$ of Ref.~\cite{Rogers_CJP1977}.

The decay properties of the IAS obtained in the present paper are shown in Table~\ref{T7902}. The sum of the intensities for the 7902- and 6289-keV $\beta$-delayed $\gamma$ rays through the IAS is measured to be $I_{\gamma}=0.20(3)\%$. The shell-model predicts a 19.3(13)\% branch for unobserved weak $\gamma$ rays, so we obtain a corrected $I_{\gamma}=0.24(3)\%$. Based on the relative proton branching ratios measured in previous $^{25}$Si decay experiments~\cite{Sextro_Thesis1973,Hatori_NPA1992,Robertson_PRC1993,Thomas_EPJA2004} and normalized to our $^{25}$Si($\beta p\gamma$)$^{24}$Mg intensities, we have determined an $I_p=13.1(16)\%$ for the IAS. Adding $I_{\gamma}$ and $I_p$ yields the total $\beta$ feeding intensity $I_\beta=13.4(16)\%$ for the IAS, corresponding to a log~$ft$ value of 3.23(6). The USDC and USDI shell-model calculations predicted the log~$ft=3.31$ and 3.39, respectively. The agreement between the measured and calculated values is good considering the theoretical uncertainties.

\begin{table}
\caption{\label{T7902}Decay properties of the 7902-keV $5/2^+$, $T=3/2$ IAS in $^{25}$Al.}
\begin{center}
\begin{ruledtabular}
\begin{tabular}{ccccc}
Reference & $\Gamma_\gamma$~(eV) & $\Gamma_p$~(eV) & $\Gamma_\gamma/\Gamma_p$ & $\omega\gamma$~(eV) \\
\hline
Ref.~\cite{Morrison_PR1968} & 2.0(10) & 111(17)\footnotemark[1] & 0.018(9) & 1.0(5) \\
Ref.~\cite{Rogers_CJP1977} & 0.50(13) & 111(17)\footnotemark[1] & 0.0045(13) & 0.25(6) \\
USDC & 2.98 & 111 & 0.027 & 1.49 \\
USDI  & 2.45 & 111 & 0.022 & 1.22 \\
Present & 2.1(5) & 111(17)\footnotemark[1] & 0.019(3) & 1.0(4) \\
\end{tabular}%
\end{ruledtabular}
\footnotetext[1]{Weighted average of $\Gamma_p$ reported in Refs.~\cite{Ikossi_NPA1976,Ikossi_PRL1976,Wilkerson_NPA1992}.}
\end{center}
\end{table}

\begin{figure}
\begin{center}
\includegraphics[width=8.6cm]{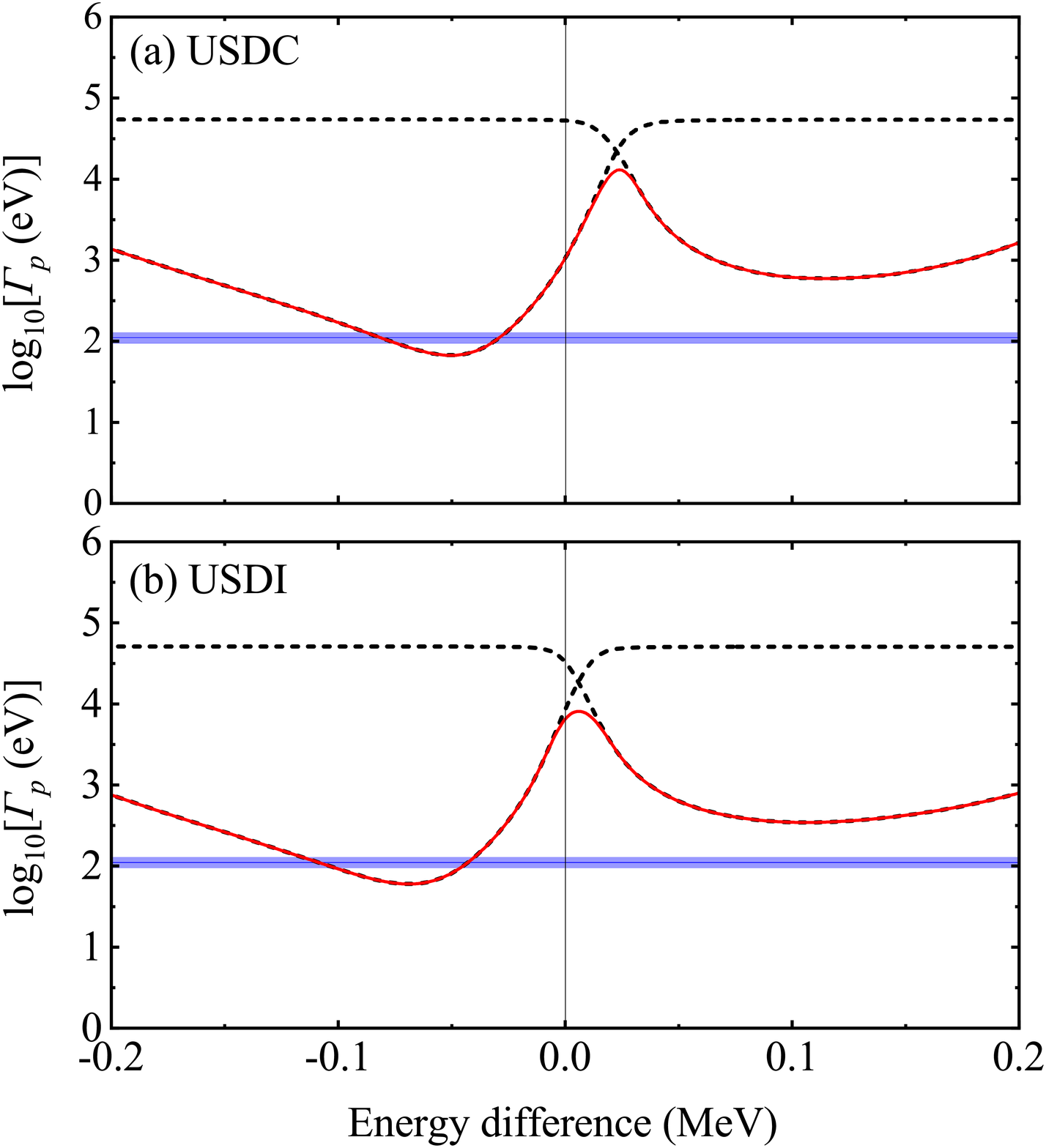}
\caption{\label{FUSD}$\Gamma_p$ for the $T=3/2$ IAS in $^{25}$Al and its neighboring $T=1/2$ state as a function of the energy difference between the two states ($E_{x3/2}-E_{x1/2}$) calculated by the (a) USDC and (b) USDI shell models. The decay width for the state dominated by the $T=3/2$ IAS is shown by the solid red line, and decay width for the state dominated by the $T=1/2$ state is shown by the dashed black line. The solid blue band corresponds to the uncertainty of the experimental $\Gamma_p$ value for the $T=3/2$ IAS~\cite{Ikossi_NPA1976,Ikossi_PRL1976,Wilkerson_NPA1992}.}
\end{center}
\end{figure}

We extract the ratio of $\Gamma_\gamma/\Gamma_p=0.019(3)$ from the $I_{\gamma}$ and $I_p$ values. Combining our $\Gamma_\gamma/\Gamma_p$ ratio with the literature $\Gamma_p$ value~\cite{Ikossi_NPA1976,Ikossi_PRL1976,Wilkerson_NPA1992} yields a $\Gamma_\gamma=2.1(5)$~eV. We have determined an $\omega\gamma=1.0(4)$~eV by taking into account the $I_{p0}/I_{p\mathrm{tot}}=0.17(5)$, deduced from the intensities of proton emission from the IAS (see Table~\ref{T25Al}). As can be seen from Table~\ref{T7902}, our results are in good agreement with the results reported by Morrison~\textit{et al}.~\cite{Morrison_PR1968} and are more precise, but they deviate from the values reported by Rogers~\textit{et al}.~\cite{Rogers_CJP1977} roughly by a factor of 4.

The USDC and USDI shell-model calculations estimate the $\Gamma_\gamma$ to be 2.98 and 2.45~eV, respectively, in agreement with the $\Gamma_\gamma=2.1(5)$~eV derived from our $\Gamma_\gamma/\Gamma_p$ ratio and the $\Gamma_p$ from Refs.~\cite{Ikossi_NPA1976,Ikossi_PRL1976,Wilkerson_NPA1992}. The shell model also indicates that $\Gamma_p$ of the IAS depends on the mixing with a predicted nearby $5/2^+$, $T=1/2$ state that has a $\Gamma_p$ of 50~keV. Unfortunately, this state has not yet been identified experimentally. The sum of Fermi and Gamow-Teller contributions is derived to be $B(\text{F})+(g_A/g_V)^2B(\text{GT})=3.6(5)$ from our measured log~$ft=3.23(6)$. Our shell-model calculations predict a $B(\text{GT})\approx0.1$ for the IAS. The summed $B$(F) should fulfill the sum rule $\sum B(\text{F})=3$, suggesting that the Fermi strength is mainly concentrated on the IAS. The fragmentation of the Fermi strength via isospin mixing is rather small compared with the strong mixing observed for some special cases~\cite{Iacob_PRC2006,Tripathi_PRL2013,Orrigo_PRL2014,Bennett_PRL2016}. The USDC and USDI shell-model calculations predict that this state is 23 and 9~keV above the IAS, respectively, but there is an uncertainty of about 150~keV for the predicted energy of each state. It has been shown that this energy uncertainty leads to uncertainties of about an order of magnitude for the proton and neutron decay width of IAS in the $sd$ shell~\cite{Ormand_PLB1986}. In order to assess the results for $^{25}$Al, we move the relative location of the $T=3/2$ and $1/2$ states by adding $b\hat{T}^2$ to the Hamiltonian, where $\hat{T}$ is the isospin operator. For states with good isospin, $b\hat{T}^2|T\rangle=bT(T+1)|T\rangle$. The $T=3/2$ states are shifted by $3.75b$, and the $T=1/2$ states are shifted by $0.75b$. The results obtained for the IAS and its neighboring $T=1/2$ state obtained by adjusting the parameter $b$ are shown in Fig.~\ref{FUSD}. If the IAS is moved down by a few keV, the $\Gamma_p$ for the IAS comes into agreement with the well-measured $\Gamma_p$ value~\cite{Ikossi_NPA1976,Ikossi_PRL1976,Wilkerson_NPA1992}. This is equivalent to moving the $T=1/2$ state up a few~keV. An empirical two-state mixing formalism~\cite{Tripathi_PRL2013} predicts that the unperturbed and observed level spacing of the two states differ by less than 1~keV in this case. The USDC and USDI shell model predicts that the $T=1/2$ state is approximately 30 and 44~keV above the $T=3/2$ IAS, respectively, corresponding to 7932$-$7946~keV for the excitation energy of the $T=1/2$ state.

\section{Conclusion}
By using the GADGET system at NSCL, simultaneous measurements of $^{25}$Si $\beta$-delayed proton and $\gamma$ decays were carried out. We have reported the most precise half-life of $^{25}$Si to date. Eight new $\beta$-delayed $\gamma$-ray transitions were detected, leading to the population of three $^{25}$Al states that have not been previously observed via $^{25}$Si $\beta$-delayed $\gamma$ decay. A total of 14 $\beta$-delayed proton branches have been identified, including a new proton peak at 724~keV. We have confirmed the placement of protons in the decay scheme of $^{25}$Si reported by previous literature~\cite{Robertson_PRC1993,Thomas_EPJA2004} using both Doppler broadening line-shape analysis and proton-$\gamma$-ray coincidence analysis. We have reevaluated the energies and intensities for 34 $^{25}$Si $\beta$-delayed proton emissions. A more precise lifetime for the $^{25}$Al 2673-keV state has been extracted, and the discrepancy involving the $\gamma$-ray partial width of the 7902-keV $T=3/2$ IAS in $^{25}$Al in the literature has been resolved, which demonstrates the potential of utilizing complementary experimental approaches. The mirror-asymmetry parameters have been deduced for five transitions in the mirror $\beta$ decays of $^{25}$Si and $^{25}$Na, which will contribute to the systematic understanding of the nature of mirror-symmetry breaking. Shell-model calculations using the USDC and USDI Hamiltonians both reproduce the experimental data well and predict a $5/2^+$, $T=1/2$ state above the $T=3/2$ IAS. It is desirable for future experiments with higher statistics to explore the fine structure near the IAS and search for the hypothetical $T=1/2$ $^{25}$Al state that exhibits weak isospin mixing with the IAS.

\begin{acknowledgments}
We gratefully acknowledge the NSCL staff for technical assistance and for providing the $^{25}$Si beam. We would like to thank Timilehin Ogunbeku and Yongchi Xiao for helpful discussions. This work was supported by the U.S. National Science Foundation under Grants No. PHY-1102511, PHY-1565546, PHY-1913554, and PHY-1811855, and the U.S. Department of Energy, Office of Science, under Award No. DE-SC0016052. L.S. acknowledges support from the Office of China Postdoctoral Council under Grant No.~20180068.
\end{acknowledgments}

\end{document}